\newcommand{\lvb}{\Bigl \bracevert}

\documentclass[pre, amssymb, showkeys, showpacs]{revtex4}

\usepackage{graphicx}
\begin{document}

\title{Solution of the Quasispecies Model for an Arbitrary Gene Network}

\author{Emmanuel Tannenbaum}
\email{etannenb@fas.harvard.edu}
\author{Eugene I. Shakhnovich}
\affiliation{Department of Chemistry and Chemical Biology,
Harvard University, Cambridge, MA 02138}

\begin{abstract}

In this paper, we study the equilibrium behavior of Eigen's
quasispecies equations for an arbitrary gene network.  We consider
a genome consisting of $ N $ genes, so that each gene sequence $
\sigma $ may be written as $ \sigma = \sigma_1 \sigma_2 \dots
\sigma_N $.  We assume a single fitness peak (SFP) model for each
gene, so that gene $ i $ has some ``master'' sequence $ \sigma_{i,
0} $ for which it is functioning.  The fitness landscape is then
determined by which genes in the genome are functioning, and which
are not.  The equilibrium behavior of this model may be solved in
the limit of infinite sequence length.  The central result is
that, instead of a single error catastrophe, the model exhibits a
series of localization to delocalization transitions, which we
term an ``error cascade.'' As the mutation rate is increased, the
selective advantage for maintaining functional copies of certain
genes in the network disappears, and the population distribution
delocalizes over the corresponding sequence spaces. The network
goes through a series of such transitions, as more and more genes
become inactivated, until eventually delocalization occurs over
the entire genome space, resulting in a final error catastrophe.
This model provides a criterion for determining the conditions
under which certain genes in a genome will lose functionality due
to genetic drift.  It also provides insight into the response of
gene networks to mutagens. In particular, it suggests an approach
for determining the relative importance of various genes to the
fitness of an organism, in a more accurate manner than the
standard ``deletion set'' method. The results in this paper also
have implications for mutational robustness and what C.O. Wilke
termed ``survival of the flattest.''

\end{abstract}



\maketitle

\section{Introduction}

A challenging problem in quantitative biology is to successfully model
the evolutionary response of organisms to various environmental pressures.
Aside from its intrinsic interest, the development of models which can
predict the time evolution of a population's genotype could prove useful
in understanding a number of important phenomena, such as antibiotic drug
resistance, cancer, viral replication dynamics, and immune response.

Perhaps the simplest formalism for modeling, at least phenomenologically,
the evolutionary dynamics of replicating organisms is known as the
quasispecies model \cite{EIG1, EIG2}.  This model was introduced by Manfred
Eigen in 1971 as a way to describe the {\it in vitro} evolution of
single-stranded RNA genomes \cite{EIG1}.  In the simplest formulation of the
model, we consider a population of asexually replicating genomes, whose only
source of variability is induced by point mutations during replication.
We assume that each genome, denoted by $ \sigma $, may be written as
$ \sigma = s_1 \dots s_L $, where each ``base'' $ s_i $ is drawn from
an alphabet of size $ S $.  With each genome is associated a first-order
growth rate constant $ \kappa_{\sigma} $,  which we assume to be
genome-dependent, since different genomes are expected to be differently
suited to the given environment.  The set of all growth rate
constants is termed the {\it fitness landscape}, which will generally be
time-dependent.

Replication and mutation give rise to mutational flow between the genomes.
If we let $ n_{\sigma} $ denote the number of organisms with genome $ \sigma $,
then,
\begin{equation}
\frac{d n_{\sigma}}{dt} = \sum_{\sigma'}{\kappa_m(\sigma', \sigma) n_{\sigma'}}
\end{equation}
where $ \kappa_m(\sigma', \sigma) $ denotes the first-order mutation rate
constant from $ \sigma' $ to $ \sigma $.  If $ p_m(\sigma', \sigma) $
denotes the probability that, after replication, $ \sigma' $ produces the
daughter genome $ \sigma $, then clearly $ \kappa_m(\sigma', \sigma) =
\kappa_{\sigma'} p_m(\sigma', \sigma) $.  To compute $ p_m(\sigma', \sigma) $,
we assume a per base replication error probability $ \epsilon_{\sigma} $ for
genome $ \sigma $ (different genomes may have different replication error
probabilities, since some genomes may code for various repair mechanisms which
other genomes do not).  It is then readily shown that \cite{REPFULL},
\begin{equation}
p_m(\sigma', \sigma) = (\frac{\epsilon_{\sigma'}}{S-1})^{D_H(\sigma, \sigma')}
(1 - \epsilon_{\sigma'})^{L - D_H(\sigma, \sigma')}
\end{equation}
where $ D_H(\sigma, \sigma') $ denotes the Hamming distance between
$ \sigma $ and $ \sigma' $.

In order to model the relative competition between various genomes, it
proves convenient to reexpress the dynamics in terms of population fractions.
Defining $ n = \sum_{\sigma}{n_{\sigma}} $, and $ x_{\sigma} =
n_{\sigma}/n $, we obtain the system of equations,
\begin{equation}
\frac{d x_{\sigma}}{dt} = \sum_{\sigma'}{\kappa_m(\sigma', \sigma) x_{\sigma'}}
- \bar{\kappa}(t) x_{\sigma}
\end{equation}
where $ \bar{\kappa}(t) \equiv \sum_{\sigma}{\kappa_{\sigma} x_{\sigma}} $,
and is therefore simply the mean fitness of the population.

The above system of equations is physically realizable in a
chemostat, which continuously siphons off organisms to maintain a
constant population size \cite{CHEMOSTAT}.  This ensures that
growth is not resource limited, so the assumption of simple
exponential growth is a good one.  It should be pointed out,
however, that it is possible to introduce a death term which
places a cap on the population size, without changing the form of
the quasispecies equations.  If we introduce a second-order
crowding term (logistic growth), so that,
\begin{equation}
\frac{d n_{\sigma}}{dt} = \sum_{\sigma'}{\kappa_m(\sigma', \sigma) n_{\sigma'}}
- k_d n_{\sigma} n
\end{equation}
then if $ k_d $ is genome-independent, it is readily shown that when converting
to the $ x_{\sigma} $ the quasispecies equations are unchanged.

The quasispecies equations may be written in vector form as,
\begin{equation}
\frac{d \vec{x}}{dt} = {\bf A}\vec{x} - (\vec{\kappa} \cdot \vec{x}) \vec{x}
\end{equation}
where $ \vec{x} = (x_{\sigma}) $ is the vector of population
fractions, $ {\bf A} = ((A_{\sigma \sigma'} = \kappa_m(\sigma',
\sigma))) $ is the matrix of first-order mutation rate constants,
and $ \vec{\kappa} = (\kappa_{\sigma}) $ is the vector of
first-order growth rate constants. For a static fitness landscape,
Eigen proved that $ \vec{x} $ evolves to the equilibrium
distribution given by the eigenvector corresponding to the largest
eigenvalue of $ {\bf A} $ \cite{EIG1, EIG2, GALLUCCIO}.

A considerable amount of research on quasispecies theory has
focused on the simplest possible fitness landscape, known as the
{\it single fitness peak} (SFP) landscape \cite{GALLUCCIO,
TARAZONA, SATORRAS, PELITI, SWETINA, ALTMEYER, NILSNOAD1,
NILSNOAD2}  .  In the SFP model, there exists a single, ``master''
sequence $ \sigma_0 $ for which $ \kappa_{\sigma_0} = k > 1 $,
while for all other sequences we have $ \kappa_{\sigma} = 1 $.
The SFP model assumes a genome-independent mutation rate, so that
$ \epsilon_{\sigma} = \epsilon $ for all $ \sigma $.

The SFP landscape is analytically solvable in the limit of
infinite sequence length.  The equilibrium behavior of the model
exhibits two distinct regimes:  A localized regime, where the
genome population clusters about the master sequence (giving rise
to the term ``quasispecies''), and a delocalized regime, where the
genome population is distributed essentially uniformly over the
entire sequence space.  The transition between the two regimes is
known as the {\it error catastrophe}, and can be shown to occur
when $ p_{rep} $, the probability of correctly replicating a
genome, drops below $ 1/k $ \cite{GALLUCCIO}.  The error
catastrophe is generally regarded as the central result of
quasispecies theory, and it has been experimentally verified in
both viruses \cite{CAT1} and bacteria \cite{CAT2}. Indeed, the
error catastrophe has been shown to be the basis for a number of
anti-viral therapies \cite{CAT1}.

The structure of the quasispecies equations naturally lends itself
to application to more complex systems than RNA molecules. Indeed,
the model has been used to successfully model certain aspects of
the immune response to viral infection \cite{BCELL}. However, in
their original form, the quasispecies equations fail to capture a
number of important aspects of the evolutionary dynamics of real
organisms.  For example, it is implicitly assumed that each genome
replicates {\it conservatively}, meaning that the original genome
is preserved by the replication process.  Correct modeling of
DNA-based life must take into account the fact that DNA
replication is {\it semiconservative} \cite{VOET}. Furthermore,
the assumption of a genome-independent replication error
probability is also too simple, since cells often have various
repair mechanisms which may become inactivated due to mutations
\cite{VOET}.  In addition, Eigen's model neglects the effects of
recombination, transposition, insertions, deletions, and gene
duplication, to name a few additional sources of variability.
Thus, a considerable amount of work remains to be done before a
quantitative theory of evolutionary response is developed.

Nevertheless, some progress has been made.  For example, semiconservative
replication was recently incorporated into the quasispecies model \cite{SEMICONSERV}.
A simple model incorporating genetic repair was developed in
\cite{REPFULL, REPFIRST}.  Diploidy has been studied in \cite{DIP}, and finite
size effects in \cite{FINSZ1, FINSZ2}.

One area in which more realistic models need to be developed is in the nature
of the fitness landscape.  As mentioned previously, the most common landscape
studied thus far has been the single fitness peak.  However, genomes
generally contain numerous genes (even the simplest of bacteria, the
Mycoplasmas, have several hundred genes \cite{KOONIN}), which work in
concert to confer viability to the organism.  Therefore, in this paper,
we consider the behavior of the model for an arbitrary gene network.  We
assume conservative replication and a genome-independent error rate for
simplicity, though we hypothesize at the end of the paper how our results
change for the case of semiconservative replication.

This paper is organized as follows:  In the following section, we
introduce our generalized $ N $-gene model defining the ``gene
network.''  We first give the quasispecies equations in terms of
the population fractions of each of the various genomes.  We
proceed to the infinite sequence length equations, and then obtain
a reduced system of equations which dictates the equilibrium
solution of our model.  We solve the model in Section III.  For
the sake of completeness, we include a simple example to
illustrate how our solution method may be applied to specific
systems. We go on in Section IV to discuss the results and
implications of our model, such as the relation to C.O. Wilke's
``survival of the flattest'' \cite{WILKE1, WILKE2, SCHUSTER}, and
also what our model says about the response of gene networks to
mutagens. Finally, we conclude in Section V with a summary of our
results and future research plans.

\section{The $ N $-Gene Model}

\subsection{Basic Equations}

Consider a population of conservatively replicating, asexual organisms,
whose genomes consist of $ N $ genes.  Each genome $ \sigma $ may then be
written as $ \sigma = \sigma_1 \dots \sigma_N $.  Let us assume, for
simplicity, a ``single fitness peak'' landscape for each gene.  That is, for
each gene $ i $ there is a ``master'' sequence $ \sigma_{i, 0} $ for which
the gene is functional, while for all $ \sigma_i \neq \sigma_{i, 0} $ the
gene is nonfunctional.  We assume that the fitness associated with a given
genome $ \sigma $ is dictated by which genes in the genome are functional,
and which are not.  We let $ \kappa_{\{i_1, \dots, i_n\}} $ denote the
fitness of organisms with genome $ \sigma $ such that $ \sigma_i =
\sigma_{i, 0} $ for $ i \in \{1, \dots, N\}/\{i_1, \dots, i_n\} $, while
$ \sigma_i \neq \sigma_{i, 0} $ for $ i \in \{i_1, \dots, i_n\} $
(we adopt the convention that $ \{i_1, \dots, i_n\} = \{\} = \emptyset $ when
$ n = 0 $).

The choice of the landscape $ \{\kappa_{\{i_1, \dots, i_n\}}|
\{i_1, \dots, i_n\} \subseteq \{1, \dots, N\}, n = 0, 1, \dots, N\} $ is
arbitrary, so that the activity of the various genes in the genome are
generally correlated.  Thus, the $ N $ genes may be regarded as defining a
gene network.  We assume that the fitnesses are all strictly positive.
Without loss of generality (i.e., by an appropriate rescaling of the time),
we may assume that $ \kappa_{\{1, \dots, N\}} = 1 $.

The simplest quasispecies equations for this $ N $-gene model are obtained
by assuming a genome-independent per base replication error probability
$ \epsilon $.  We assume that gene $ i $ has a sequence length $ L_i $, and we
define $ L = L_1 + \dots + L_N $.  Then $ p_m(\sigma', \sigma) =
p_m(\sigma_1', \sigma_1) \cdot \dots \cdot p_m(\sigma_N', \sigma_N) $, where,
\begin{equation}
p_m(\sigma_i', \sigma_i) = (\frac{\epsilon}{S-1})^{D_H(\sigma_i', \sigma_i)}
(1 - \epsilon)^{L_i - D_H(\sigma_i', \sigma_i)}
\end{equation}
Putting everything together, we obtain the system of equations,
\begin{eqnarray}
\frac{d x_{\sigma_1 \dots \sigma_N}}{dt} & = &
\sum_{\sigma_1'}\cdot \dots \cdot \sum_{\sigma_N'}
\kappa_{\sigma_1' \dots \sigma_N'} \prod_{i =
1}^{N}{(\frac{\epsilon}{S-1})^{D_H(\sigma_i', \sigma_i)}} \times
\nonumber \\
&   & (1 - \epsilon)^{L_i - D_H(\sigma_i', \sigma_i)} x_{\sigma_1'
\dots \sigma_N'} \nonumber \\
&   & - \bar{\kappa}(t) x_{\sigma_1 \dots \sigma_N}
\end{eqnarray}

Define the {\it Hamming class} $ C_H(l_1, \dots, l_N) = \{\sigma =
\sigma_1 \dots \sigma_N|D_H(\sigma_i, \sigma_{i,0}) = l_i, i = 1,
\dots, N\} $.  Also, define $ z_{l_1, \dots, l_N} = \sum_{\sigma
\in C_H(l_1, \dots, l_N)}{x_{\sigma}} $.  By the symmetry of the
landscape, we may assume that $ x_{\sigma} $ depends only on the $
l_i $ corresponding to $ \sigma $, and hence we may look at the
total population fraction in $ C_H(l_1, \dots, l_N) $ and study
its dynamics.  The conversion of the quasispecies equations in
terms of $ x_{\sigma} $ to $ z_{l_1, \dots, l_N} $ is accomplished
by a generalization of the method given in \cite{REPFULL}. The
result is,
\begin{widetext}
\begin{eqnarray}
\frac{d z_{l_1, \dots, l_N}}{dt} & = & \sum_{l_{1,1} = 0}^{L_1 -
l_1} \sum_{l_{1,2} = 0}^{l_1} \cdot \dots \cdot \sum_{l_{N,1} =
0}^{L_N - l_N} \sum_{l_{N,2} = 0}^{l_N} \prod_{i = 1}^{N}{{{L_i -
l_i - l_{i,1} + l_{i,2}}\choose
{l_{i,2}}}} {{l_{i,1} + l_i - l_{i,2}}\choose{l_{i,1}}}\times \nonumber \\
&   & \epsilon^{l_{i,2}} (1 - \epsilon)^{L_i - l_i - l_{i,1}}
(\frac{\epsilon}{S-1})^{l_{i,1}} (1 - \frac{\epsilon}{S-1})^{l_i -
l_{i,2}} z_{l_{1,1} + l_1 - l_{1,2}, \dots, l_{N,1} + l_N -
l_{N,2}}
\nonumber \\
&   & - \bar{\kappa}(t) z_{l_1, \dots, l_N}
\end{eqnarray}
\end{widetext}
We now let the $ L_i \rightarrow \infty $ in such a way that the $
\alpha_i \equiv L_i/L $ and $ \mu \equiv L \epsilon $ remain
fixed. We assume that the $ \alpha_i $ are all strictly positive
(allowing an $ \alpha_i $ to be $ 0 $ leads to certain
difficulties which we choose not to address in this paper).  Because
the probability of correctly replicating a genome is simply $ (1 -
\epsilon)^L \rightarrow e^{-\mu} $, fixing $ \mu $ is equivalent
to fixing the genome replication fidelity in the limit of infinite
sequence length.

In this limit, it is possible to show that, for each gene $ i $,
the only terms in Eq. (8) which survive the limiting process are
the $ l_{i,1} = 0 $ terms \cite{REPFULL}.  This is equivalent to
the statement that, in the limit of infinite sequence length,
backmutations may be neglected.  We also obtain that,
\begin{equation}
{{L_i - l_i + l_{i,2}} \choose l_{i,2}} \epsilon^{l_{i,2}}
\rightarrow \frac{1}{l_{i,2}!} (\alpha_i \mu)^{l_{i,2}}
\end{equation}
and
\begin{equation}
(1 - \epsilon)^{L_i - l_i} \rightarrow e^{-\alpha_i \mu}
\end{equation}

The final result is,
\begin{eqnarray}
\frac{d z_{l_1, \dots, l_N}}{dt} & = & e^{-\mu} \sum_{l_1' =
0}^{l_1} \cdot \dots \cdot \sum_{l_N' = 0}^{l_N} \frac{\kappa_{l_1
- l_1', \dots, l_N - l_N'}} {l_1'! \cdot \dots \cdot l_N'!} \times
\nonumber \\
&   & (\alpha_1 \mu)^{l_1'} \cdot \dots \cdot (\alpha_N
\mu)^{l_N'} z_{l_1 - l_1', \dots, l_N - l_N'} \nonumber \\
&   & - \bar{\kappa}(t) z_{l_1, \dots, l_N}
\end{eqnarray}

It should be noted that the neglect of backmutations is only valid
when one can group population fractions into Hamming classes.  In
our case, by the symmetry of the fitness landscape, the
equilibrium solution only depends on the Hamming class, and hence,
to find the equilibria, it is perfectly valid to
``pre-symmetrize'' the population distribution and study the
resulting dynamics.

Thus, when studying dynamics, it is generally not valid to neglect
backmutations.  For example, consider a single fitness peak
landscape, and suppose that a population of organisms is at its
equilibrium, clustered about the fitness peak.  If the organisms
are then mutated, so that they are shifted away from the fitness
peak, then eventually they will backmutate and reequilibrate on
the fitness peak (this situation has been observed with
prokaryotes \cite{PROK}).  If we imagine that the mutation shifts
the organism from the master genome $ \sigma_0 $ to some other
genome $ \sigma' \neq \sigma_0 $, then it is clear that the
landscape is not symmetric about $ \sigma' $, and furthermore that
the population distribution is not symmetric about $ \sigma_0 $.
Thus, Eq. (11) does not apply.  To correctly model the
reequilibration dynamics, it is necessary to consider the finite
sequence length equations, and explicitly incorporate
backmutations.

\subsection{Reduced Equations}

Because of the neglect of backmutations, Eq. (11) may in principle
be solved recursively to obtain the equilibrium distribution of
the $ z_{l_1, \dots, l_N} $ at any $ \mu $, assuming we know the
equilibrium mean fitness, denoted $ \bar{\kappa}(t = \infty) $.
The problem, of course, is that $ \bar{\kappa}(t = \infty) $ needs
to be computed.  This may be done as follows:  Given any
collection $ \{i_1, \dots, i_n\} \subseteq \{1, \dots, N\} $ of
indices, define $ \tilde{z}_{\{i_1, \dots, i_n\}} $ via,
\begin{equation}
\tilde{z}_{\{i_1, \dots, i_n\}} =
\sum_{l_{i_1} = 1}^{\infty} \cdot \dots \cdot \sum_{l_{i_n} = 1}^{\infty}
{z_{l_{i_1} {\bf e}_{i_1} + \dots + l_{i_n} {\bf e}_{i_n}}}
\end{equation}
where $ {\bf e}_{1} = (1, 0, \dots, 0) $, $ {\bf e}_{2} = (0, 1, 0, \dots, 0)
$, and so forth.  Thus, $ \tilde{z}_{\{i_1, \dots, i_n\}} $ is
simply the total fraction of the population in which
the genes of indices $ \{i_1, \dots, i_n\} $ are faulty, while
the remaining genes are given by their corresponding master
sequences.

The dynamics of the $ \tilde{z}_{\{i_1, \dots, i_n\}} $ is derived
in Appendix A.  The result is given by,
\begin{widetext}
\begin{eqnarray}
\frac{d \tilde{z}_{\{i_1, \dots, i_n\}}}{dt}
& = &
(\kappa_{\{i_1, \dots, i_n\}}
e^{-(1 - \alpha_{i_1} - \dots - \alpha_{i_n})\mu} - \bar{\kappa}(t))
\tilde{z}_{\{i_1, \dots, i_n\}} + \nonumber \\
&   & e^{-(1 - \alpha_{i_1} - \dots - \alpha_{i_n}) \mu} \sum_{k =
0}^{n-1} \sum_{\{j_1, \dots, j_k\} \subset \{i_1, \dots, i_n\}}
{\kappa_{\{j_1, \dots, j_k\}} \tilde{z}_{\{j_1, \dots, j_k\}}
\prod_{i \in \{i_1, \dots, i_n\}/\{j_1, \dots, j_k\}}
(1 - e^{-\alpha_i \mu})} \nonumber \\
\end{eqnarray}
\end{widetext}
We can provide an intuitive explanation for this expression:
Because backmutations may be neglected in the limit of infinite
sequence length, it follows that, once a gene is disabled, it
remains disabled. Therefore, given a set of indices $ \{i_1,
\dots, i_n\} $, mutational flow can only occur from $
\tilde{z}_{\{i_1, \dots, i_n\}} $ to $ \tilde{z}_{\{j_1, \dots,
j_m\}} $ for which $ \{i_1, \dots, i_n\} \subseteq \{j_1, \dots,
j_m\} $ (in this paper, if $ \Omega_1 \subset \Omega_2 $, then $
\Omega_1 $ is a proper subset of $ \Omega_2 $.  If $ \Omega_1
\subseteq \Omega_2 $, then either $ \Omega_1 $ is a proper subset
of $ \Omega_2 $ or $ \Omega_1 = \Omega_2 $).  Similarly, $
\tilde{z}_{\{i_1, \dots, i_n\}} $ can only receive mutational
contributions from $ \tilde{z}_{\{j_1, \dots, j_m\}} $ for which $
\{j_1, \dots, j_m\} \subseteq \{i_1, \dots, i_n\} $. For such a $
\{j_1, \dots, j_m\} $, the probability of mutation to $ \{i_1,
\dots, i_n\} $ may be computed as follows: Because the genes
corresponding to the indices $ j_1, \dots, j_m $ remain faulty,
the neglect of backmutations means that it does not matter whether
these genes are replicated correctly or not.  All genes with
indices in $ \{1, \dots, N\}/\{i_1, \dots, i_n\} $ must remain
equal to the corresponding master sequences after mutation. The
probability that gene $ i $ replicates correctly is given by $
e^{-\alpha_i \mu} $, so the probability that all genes with
indices in $ \{1, \dots, N\}/\{i_1, \dots, i_n\} $ replicate
correctly is $ \prod_{i \in \{1, \dots, N\}/\{i_1, \dots, i_n\}}
{e^{-\alpha_i \mu}} = e^{-(1 - \alpha_{i_1} - \dots -
\alpha_{i_n})\mu} $. The genes which must be replicated
incorrectly are those with indices in $ \{i_1, \dots, i_n\}/\{j_1,
\dots, j_m\} $.  Since each such gene replicates incorrectly with
probability $ 1 - e^{-\alpha_i \mu} $, it follows that the
probability of replicating all genes in $ \{i_1, \dots,
i_n\}/\{j_1, \dots, j_m\} $ incorrectly is $ \prod_{i \in \{i_1,
\dots, i_n\}/\{j_1, \dots, j_m\}}{(1 - e^{-\alpha_i \mu})} $.
Putting everything together, we obtain a mutational flow from $
\tilde{z}_{\{j_1, \dots, j_m\}} $ to $ \tilde{z}_{\{i_1, \dots,
i_n\}} $ of $ e^{-(1 - \alpha_{i_1} - \dots - \alpha_{i_n})\mu}
\kappa_{\{j_1, \dots, j_m\}} \tilde{z}_{\{j_1, \dots, j_m\}}
\prod_{i \in \{i_1, \dots, i_n\}/\{j_1, \dots, j_m\}} {(1 -
e^{-\alpha_i \mu})} $.  Summing over all possible $ \{j_1, \dots,
j_m\} \subseteq \{i_1, \dots, i_n\} $ gives us the expression in
Eq. (13).

Note that $ \bar{\kappa}(t) = \sum_{n = 0}^{N} \sum_{\{i_1, \dots,
i_n\}} {\kappa_{\{i_1, \dots, i_n\}} \tilde{z}_{\{i_1, \dots,
i_n\}}} $, so we need to solve Eq. (13) in order to obtain the
equilibrium distribution of the model.

\section{Solution of the Model}

In this section, we proceed to solve the reduced system of
equations given by Eq. (13).  Since this provides us with $
\bar{\kappa}(t = \infty) $ and $ z_{0, \dots, 0} =
\tilde{z}_{\emptyset} $, it follows that we can recursively solve
for the equilibrium values of all $ z_{l_1, \dots, l_N} $.

In vector notation, Eq. (13) may be expressed in the form,
\begin{equation}
\frac{d \vec{\tilde{z}}}{dt} = {\bf B} \vec{\tilde{z}} -
(\vec{\kappa} \cdot \vec{\tilde{z}}) \vec{\tilde{z}}
\end{equation}
where $ \vec{\tilde{z}} $ is the vector of all $ \tilde{z}_{\{i_1,
\dots, i_n\}} $, $ \vec{\kappa} $ is the vector of all $
\kappa_{\{i_1, \dots, i_n\}} $, and $ {\bf B} $ is the matrix of
mutation rate constants.

Because of the neglect of backmutations in the limit of infinite
sequence length, different regions of the genome space become
mutationally decoupled, so that the largest eigenvalue of the
mutation matrix $ {\bf B} $ will in general be degenerate.  Thus,
the equilibrium of the reduced system of equations is not unique.
However, for any initial condition, the system will evolve to an
equilibrium, though of course different initial conditions will
yield different equilibrium results.

\subsection{Definitions}

In this subsection, we define a variety of constructs which we will
need to characterize the equilibrium behavior of our model.  We begin
with the definition of a {\it node}:  We define a {\it level n node} to refer
to any collection of ``knocked out'' genes with indices
$ \{i_1, \dots, i_n\} \subseteq \{1, \dots, N\} $.  The reason for this
terminology is simple.  We may imagine the set of all nodes to be connected
via mutations.  Because of the neglect of backmutations, it follows
that a node $ \{i_1, \dots, i_n\} $ is accessible from a node
$ \{j_1, \dots, j_m\} $ via mutations if and only if
$ \{j_1, \dots, j_m\} \subseteq \{i_1, \dots, i_n\} $.  The result is that
we can generate a directed graph of mutational flows between nodes, an
example of which is illustrated in Figure 1.

\begin{figure}
\includegraphics[width = 0.9\linewidth]{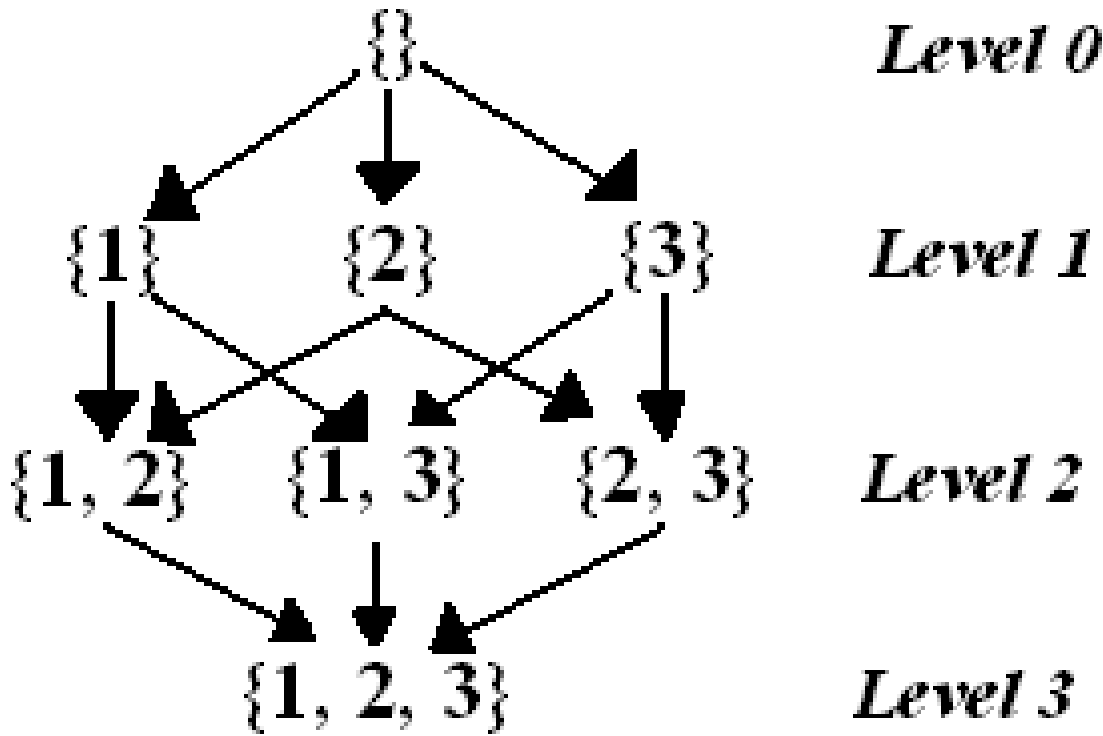}
\caption{The directed graph of mutational flow between nodes for a
three-gene network.}
\end{figure}

Given some node $ \nu = \{i_1, \dots, i_n\} $, define
$ G_\nu \equiv \{\tilde{\nu} \subseteq \{1, \dots, N\}|\nu
\subseteq \tilde{\nu}\} $.  Therefore, $ G_\nu $ may be regarded as the
subgraph of all nodes which are mutationally accessible from
$ \nu $.  An example of such a subgraph is illustrated
in Figure 2.

\begin{figure}
\includegraphics[width = 0.9\linewidth]{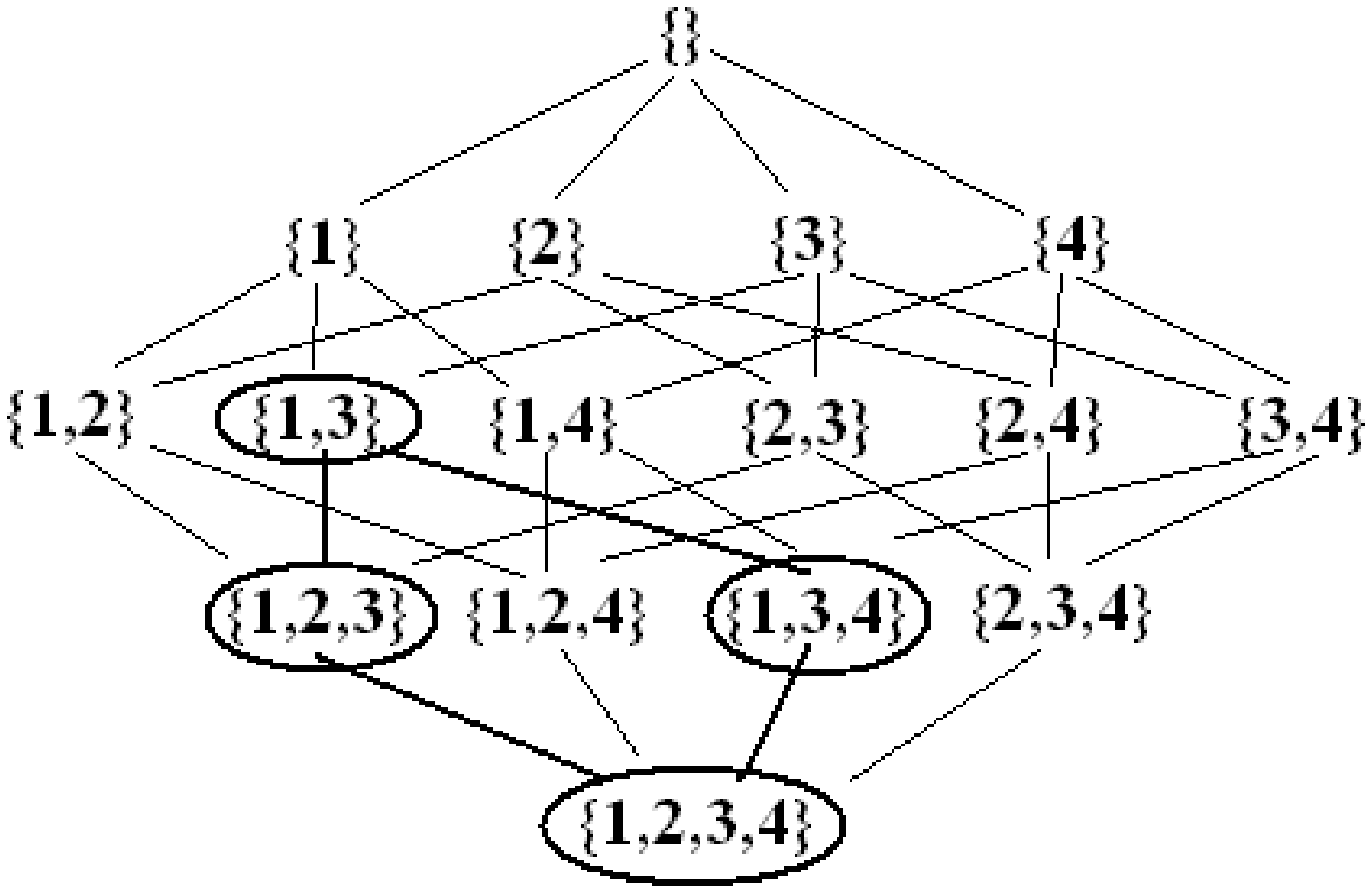}
\caption{The mutational subgraph $ G_{\{1,3\}} $ for a four-gene
network.}
\end{figure}

Let $ \Omega $ denote any collection of nodes.  Then we may define
$ G_{\Omega} \equiv \bigcup_{\nu \in \Omega}{G_\nu} $.  Furthermore,
define $ \tilde{\Omega} = \{\nu \in \Omega|\Omega \cap G_\nu = \nu\} $.
Thus, $ \tilde{\Omega} $ is the set of all nodes in $ \Omega $ such that
no node in $ \Omega $ is contained within the mutational subgraph
of any other node in $ \tilde{\Omega} $.  Figure 3 gives an example
showing the construction of $ \tilde{\Omega} $ from $ \Omega $.

\begin{figure}
\includegraphics[width = 0.9\linewidth]{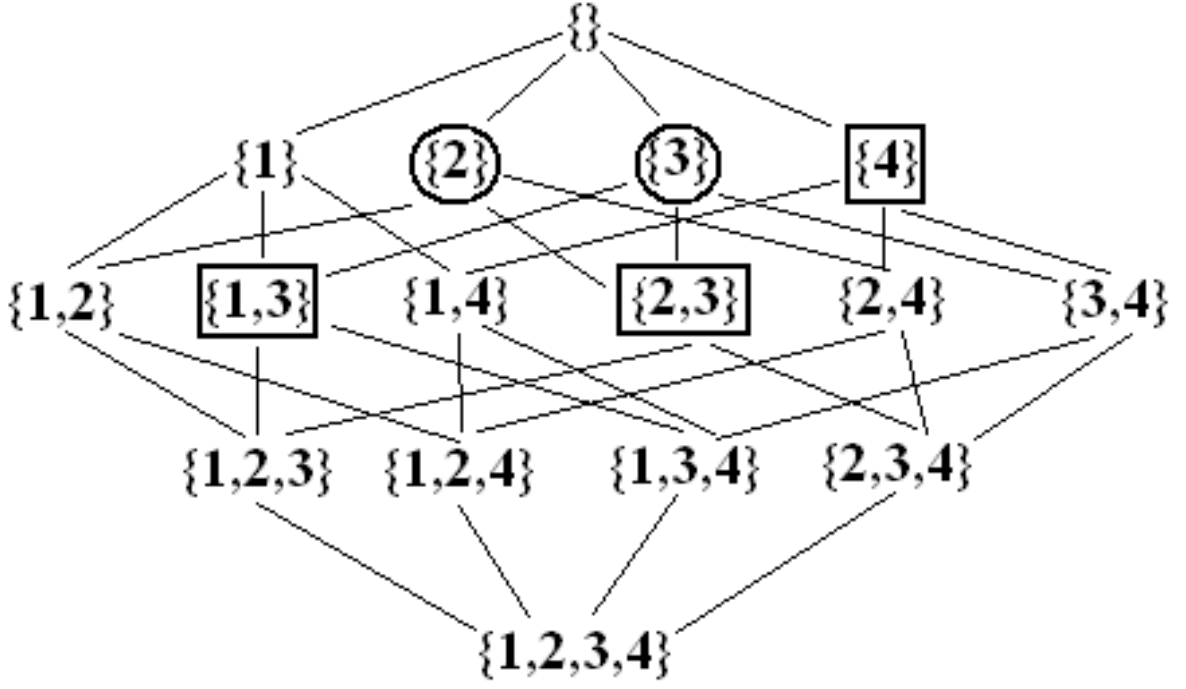}
\caption{Illustration of $ \Omega $ and $ \tilde{\Omega} $ in
a four-gene network.  The nodes circled with rectangles and circles
constitute $ \Omega $.  The nodes circled only with rectangles
constitute $ \tilde{\Omega} $.}
\end{figure}

Given some node $ \{i_1, \dots, i_n\} $, define $
\kappa_{eff}(\{i_1, \dots, i_n\}; \mu) = \kappa_{\{i_1, \dots,
i_n\}} e^{-(1 - \alpha_{i_1} - \dots - \alpha_{i_n})\mu} $. We
then define $ \kappa_{max}(\mu) = max\{\kappa_{eff}(\nu; \mu)|\nu
\subseteq \{1, \dots, N\}\} $.  Finally, given some $ \mu $,
define $ \Omega_{max}(\mu) = \{\nu \subseteq \{1, \dots, N\}|
\kappa_{eff}(\nu; \mu) = \kappa_{max}(\mu)\} $.

With these definitions in hand, we are now ready to obtain the structure
of the equilibrium solution at a given $ \mu $.

\subsection{Equilibrium Solution}

\subsubsection{Determination of $ \bar{\kappa}(t = \infty) $}

We claim that $ \bar{\kappa}(t = \infty) = \kappa_{max}(\mu) $.  We
prove this in two steps.  First of all, we claim that
$ \bar{\kappa}(t = \infty) = \kappa_{eff}(\nu; \mu) $ for some node
$ \nu $.  Clearly, because $ \sum_{\nu \subseteq \{1, \dots, N\}}
{\tilde{z}_\nu} = 1 $, it follows that at least one of the
$ \tilde{z}_\nu > 0 $ at equilibrium.  Let $ \nu' = \{i_1, \dots, i_n\} $
be a node of minimal $ n $ such that $ \tilde{z}_{\nu'} > 0 $.  Then it
should be clear that, at equilibrium, we have,
\begin{equation}
0 = \frac{d \tilde{z}_{\nu'}}{dt}\lvb_{t = \infty} = (\kappa_{eff}
(\nu'; \mu) - \bar{\kappa}(t = \infty)) \tilde{z}_{\nu'}
\end{equation}
which, since $ \tilde{z}_{\nu'} > 0 $, may be solved to give
$ \bar{\kappa}(t = \infty) = \kappa_{eff}(\nu'; \mu) $.

So now suppose that $ \bar{\kappa}(t = \infty) \neq
\kappa_{max}(\mu) $.  Then $ \bar{\kappa}(t = \infty) <
\kappa_{max}(\mu) $.  Such an equilibrium can never be observed
because it is unstable.  To see this, let $ \nu_{max} $ denote a
node for which $ \kappa_{eff}(\nu_{max}; \mu) = \kappa_{max}(\mu)
$.  Then from Eq. (13) we have, at equilibrium, that,
\begin{equation}
0 = \frac{d \tilde{z}_{\nu_{max}}}{dt}\lvb_{t = \infty} \geq
(\kappa_{eff}(\nu_{max}; \mu) - \bar{\kappa}(t = \infty))
\tilde{z}_{\nu_{max}}
\end{equation}
and so $ \tilde{z}_{\nu_{max}} = 0 $.  Clearly, however, any
perturbation on $ \tilde{z}_{\nu_{max}} $ will push $
\tilde{z}_{\nu_{max}} $ away from its equilibrium value.  This
equilibrium is therefore unstable, and hence, unobservable.

Note that since $ \bar{\kappa}(t = \infty) = \kappa_{max}(\mu) $,
it follows that the mean equilibrium fitness is a continuous
function of $ \mu $.

\subsubsection{Determining the $ \tilde{z}_{\{i_1, \dots, i_n\}} $}

To find the equilibrium solution of the reduced system of
equations, we first need to determine which $ \tilde{z}_{\nu} = 0
$ at equilibrium.  To this end, we begin with the claim that, for
$ \mu > 0 $, $ \tilde{z}_{\nu} = 0 $ unless $ \nu \in
G_{\tilde{\Omega}_{max}(\mu)} $.  For suppose there exists $ \nu
\notin G_{\tilde{\Omega}_{max}(\mu)} $ such that $ \tilde{z}_{\nu}
\neq 0 $ at equilibrium.  Then out of the set of all nodes which
satisfy the above two properties, we may choose $ \nu $ to be of
minimal level.  We claim that, for any $ \tilde{\nu} \subseteq \nu
$, we have that $ \tilde{\nu} \notin G_{\tilde{\Omega}_{max}(\mu)}
$, for otherwise it is clear that $ \nu \in G_{\tilde{\nu}}
\subseteq G_{\tilde{\Omega}_{max}(\mu)} \Rightarrow\Leftarrow $.
Therefore, by the minimality of the level of $ \nu $, it follows
that $ \tilde{z}_{\tilde{\nu}} = 0 $ whenever $ \tilde{\nu} $ is a
proper subset of $ \nu $.  But then the equilibrium equation for $
\tilde{z}_{\nu} $ gives $ \bar{\kappa}(t = \infty) =
\kappa_{eff}(\nu; \mu) $, and so $ \kappa_{eff}(\nu; \mu) =
\kappa_{max}(\mu) $.  Therefore, $ \nu \in \Omega_{max}(\mu) $.
However, by assumption, $ \nu \notin \tilde{\Omega}_{max}(\mu) $,
which means that $ G_{\nu} $ contains nodes in $ \Omega_{max}(\mu)
$ which are distinct from $ \nu $.  Denote one of these nodes by $
\tilde{\nu} = \{j_1, \dots, j_m\} $.  Then at equilibrium we have,
from Eq. (13), that,
\begin{eqnarray}
\frac{d \tilde{z}_{\tilde{\nu}}}{dt}\lvb_{t = \infty} & = &
(\kappa_{max}(\mu) - \kappa_{max}(\mu))\tilde{z}_{\tilde{\nu}} +
\nonumber \\
&   & e^{-(1 - \alpha_{j_1} - \dots - \alpha_{j_m})\mu} \sum_{k =
0}^{m-1} \sum_{\nu' \subset \tilde{\nu}} \times \nonumber
\\
&   & \kappa_{\nu'} \tilde{z}_{\nu'} \prod_{i \in
\tilde{\nu}/\nu'}{(1 - e^{-\alpha_i \mu})}
\nonumber \\
& \geq & e^{-(1 - \alpha_{j_1} - \dots - \alpha_{j_m})\mu}
\kappa_{\nu} \tilde{z}_{\nu} \prod_{i \in \tilde{\nu}/\nu}{(1 -
e^{-\alpha_i \mu})}
\nonumber \\
& > & 0
\end{eqnarray}
which is clearly a contradiction.  This establishes our claim.

We now argue that our equilibrium solution may be found if we know
$ \tilde{z}_{\nu} $ for $ \nu \in \tilde{\Omega}_{max}(\mu) $.
We claim that for any $ \nu \in G_{\tilde{\Omega}_{max}(\mu)} $, we may
write,
\begin{equation}
\tilde{z}_{\nu} = \sum_{\tilde{\nu} \in \tilde{\Omega}_{max}(\mu)}
{\beta_{\tilde{\nu} \nu}(\mu) \tilde{z}_{\tilde{\nu}}}
\end{equation}
where the $ \beta_{\tilde{\nu} \nu} \geq 0 $, and for $ \mu > 0 $ a
given $ \beta_{\tilde{\nu} \nu} $ is strictly positive if and only if
$ \nu \in G_{\tilde{\nu}} $.  The above expression then holds for all
$ \nu $, since we simply take $ \beta_{\tilde{\nu} \nu} = 0 $ for
$ \nu \notin G_{\tilde{\Omega}_{max}(\mu)} $.

We can prove the above formula via induction on the level of the
nodes in $ G_{\tilde{\Omega}_{max}(\mu)} $.  In doing so, we will
essentially develop an algorithm for constructing the $
\beta_{\tilde{\nu} \nu} $.  So, let us start with $ n_{min} $, the
minimal level nodes $ G_{\tilde{\Omega}_{max}(\mu)} $.  Then
clearly $ \nu \in \tilde{\Omega}_{max}(\mu) $, so that $
\beta_{\tilde{\nu}\nu} = \delta_{\tilde{\nu} \nu} $, hence the
formula is correct for $ n_{min} $.  So now suppose that, for some
$ n \geq n_{min} $, the formula is correct for all $ m $ such that
$ n_{min} \leq m \leq n $.  Then for a level $ n + 1 $ node in $
G_{\tilde{\Omega}_{max}(\mu)} $, denoted by $ \{i_1, \dots,
i_{n+1}\} $, we have, at equilibrium, that,
\begin{eqnarray}
0
& = &
(\kappa_{eff}(\{i_1, \dots, i_{n+1}\}; \mu) - \kappa_{max}(\mu))
\tilde{z}_{\{i_1, \dots, i_{n+1}\}} \nonumber \\
&   & + e^{-(1 - \alpha_{i_1} - \dots - \alpha_{i_{n+1}})\mu}
\sum_{k = 0}^{n} \sum_{\{j_1, \dots, j_k\} \subset \{i_1, \dots,
i_{n+1}\}} \times \nonumber \\
&   & \kappa_{\{j_1, \dots, j_k\}} \tilde{z}_{\{j_1, \dots, j_k\}}
\prod_{i \in \{i_1, \dots, i_{n+1}\}/\{j_1, \dots, j_k\}}
(1 - e^{-\alpha_i \mu}) \nonumber \\
& = & (\kappa_{eff}(\{i_1, \dots, i_{n+1}\}; \mu) -
\kappa_{max}(\mu))
\tilde{z}_{\{i_1, \dots, i_{n+1}\}} \nonumber \\
&   & + e^{-(1 - \alpha_{i_1} - \dots - \alpha_{i_{n+1}})\mu}
\times \nonumber \\
&   & \sum_{\nu \subset \{i_1, \dots, i_{n+1}\},
      \nu \in G_{\tilde{\Omega}_{max}(\mu)}}
\kappa_{\nu} \sum_{\tilde{\nu} \in \tilde{\Omega}_{max}(\mu)}
\times \nonumber \\
&   & \beta_{\tilde{\nu} \nu} \tilde{z}_{\tilde{\nu}} \prod_{i \in
\{i_1, \dots, i_{n+1}\}/\nu}
(1 - e^{-\alpha_i \mu}) \nonumber \\
& = &
(\kappa_{eff}(\{i_1, \dots, i_{n+1}\}; \mu) - \kappa_{max}(\mu))
\tilde{z}_{\{i_1, \dots, i_{n+1}\}} \nonumber \\
&   & + e^{-(1 - \alpha_{i_1} - \dots - \alpha_{i_{n+1}})\mu}
\times \nonumber \\
&   & \sum_{\tilde{\nu} \in \tilde{\Omega}_{max}(\mu)}
      \tilde{z}_{\tilde{\nu}} \sum_{\nu \subset \{i_1, \dots, i_{n+1}\},
      \nu \in G_{\tilde{\nu}}} \times \nonumber \\
&   & \beta_{\tilde{\nu} \nu} \kappa_{\nu} \prod_{i \in \{i_1,
\dots, i_{n+1}\}/\nu} (1 - e^{-\alpha_i \mu})
\end{eqnarray}
Now, if $ \{i_1, \dots, i_{n+1}\} \in \tilde{\Omega}_{max}(\mu) $,
then $ \beta_{\tilde{\nu} \{i_1, \dots, i_{n+1}\}} =
\delta_{\tilde{\nu} \{i_1, \dots, i_{n+1}\}} $.  Otherwise,
$ \kappa_{eff}(\{i_1, \dots, i_{n+1}; \mu) < \kappa_{max}(\mu) $,
so the equilibrium equation may be solved to give,
\begin{eqnarray}
\beta_{\tilde{\nu} \{i_1, \dots, i_{n+1}\}} & = &
\frac{e^{-(1 - \alpha_{i_1} - \dots - \alpha_{i_{n+1}})\mu}}
     {\kappa_{max}(\mu) - \kappa_{eff}(\{i_1, \dots, i_{n+1}\};
     \mu)} \times \nonumber \\
&   & \sum_{\nu \subset \{i_1, \dots, i_{n+1}\},
      \nu \in G_{\tilde{\nu}}}
\beta_{\tilde{\nu} \nu} \kappa_{\nu} \times \nonumber \\
&   & \prod_{i \in \{i_1, \dots, i_{n+1}\}/\nu}
(1 - e^{-\alpha_i \mu}) \nonumber \\
\end{eqnarray}
Note that $ \beta_{\tilde{\nu} \{i_1, \dots, i_{n+1}\}} \geq 0 $.
Furthermore, if $ \{i_1, \dots, i_{n+1}\} \notin G_{\tilde{\nu}}
$, then no proper subset of $ \{i_1, \dots, i_{n+1}\} $ is in $
G_{\tilde{\nu}} $.  Therefore, $ \{\nu \subset \{i_1, \dots,
i_{n+1}\}| \nu \in G_{\tilde{\nu}}\} = \emptyset $, so $
\beta_{\tilde{\nu} \{i_1, \dots, i_{n+1}\}} = 0 $.  Conversely, if
$ \{i_1, \dots, i_{n+1}\} \in G_{\tilde{\nu}} $, then since $
\{i_1, \dots, i_{n+1}\} \neq \tilde{\nu} $, it follows that $
\{\nu \subset \{i_1, \dots, i_{n+1}\}| \nu \in G_{\tilde{\nu}}\}
\neq \emptyset $.  Therefore, the sum in Eq. (20) is nonempty,
hence, since the $ \beta_{\tilde{\nu} \nu} $ appearing in the sum
are all strictly positive, it follows that $ \beta_{\tilde{\nu}
\{i_1, \dots, i_{n+1}\}} > 0 $.  This implies that $
\beta_{\tilde{\nu} \{i_1, \dots, i_{n+1}\}} $ is strictly positive
if and only if $ \{i_1, \dots, i_{n+1}\} \in G_{\tilde{\nu}} $,
which completes the induction step, and proves the claim.

For each $ \tilde{\nu} \in \tilde{\Omega}_{max}(\mu) $, we can
define $ \pi_{\tilde{\nu}} = \sum_{\nu \in G_{\tilde{\nu}}}
{\beta_{\tilde{\nu} \nu}} $, and then define $ \gamma_{\tilde{\nu}
\nu} = \beta_{\tilde{\nu} \nu}/\pi_{\tilde{\nu}} $ and $
w_{\tilde{\nu}} = \pi_{\tilde{\nu} \nu} \tilde{z}_{\tilde{\nu}} $.
If, for each $ \tilde{\nu} $ we also define $
\vec{\gamma}_{\tilde{\nu}} = (\gamma_{\tilde{\nu} \nu}) $, that
is, the vector of all $ \gamma_{\tilde{\nu} \nu} $, and if we
define $ \vec{\tilde{z}} = (\tilde{z}_{\nu}) $, then we obtain,
\begin{equation}
\vec{\tilde{z}} = \sum_{\tilde{\nu} \in \tilde{\Omega}_{max}(\mu)}
                       {w_{\tilde{\nu}} \vec{\gamma}_{\tilde{\nu}}}
\end{equation}
where $ \sum_{\tilde{\nu} \in \tilde{\Omega}_{max}(\mu)}
{w_{\tilde{\nu}}} = 1 $.

Note that the $ \vec{\gamma}_{\tilde{\nu}} $ form a linearly
independent set of vectors.  Therefore, if
$ \tilde{\Omega}_{max}(\mu) $ contains more than one node,
then the equilibrium solution of the reduced system of
equations is not unique, but rather is defined by the
parallelipiped $ \{\sum_{\tilde{\nu} \in
\tilde{\Omega}_{max}(\mu)}{w_{\tilde{\nu}}
\vec{\gamma}_{\tilde{\nu}}}| \sum_{\tilde{\nu} \in
\tilde{\Omega}_{max}(\mu)}{w_{\tilde{\nu}}} = 1,
w_{\tilde{\nu}} \geq 0\} $.

As mentioned earlier, the degeneracy in the equilibrium behavior
follows from the neglect of backmutations in the limit of infinite
sequence length.  The various nodes in $ \tilde{\Omega}_{max}(\mu)
$ become mutationally decoupled in this limit, which can cause the
largest eigenvalue of the mutation matrix $ {\bf B} $ to be
degenerate.  However, for {\it finite} sequence lengths, the
quasispecies dynamics will always converge to a unique solution.
In particular, if we start with the initial condition $
z_\emptyset = 1 $, then for finite sequence lengths we will
converge to the unique equilibrium solution.  Because all nodes
are mutationally connected in the infinite sequence length limit
with this initial condition, we make the assumption that the way
to find the infinite sequence length equilibrium which is the
limit of the finite sequence length equilibria is to find the
infinite sequence length equilibrium starting from the initial
condition $ z_\emptyset = 1 $.  This allows us to break the
eigenstate degeneracy in a canonical manner.

In the appendices, we describe a fixed-point iteration approach
for finding the equilibrium solution of the model.  Within this
algorithm, we also use the initial condition $ z_\emptyset = 1 $
as the analogous approach to the one above for finding the
infinite sequence length equilibrium which is the limit of the
finite sequence length equilibria.

Finally, the treatment thus far has been finding the equilibrium
solution of the reduced system of equations for $ \mu > 0 $.  The
equilibrium solution for $ \mu = 0 $ is obtained by taking the
limit of the $ \mu > 0 $ solutions, so that $ \vec{\tilde{z}}(\mu
= 0) = \lim_{\mu \rightarrow 0^+}{\vec{\tilde{z}}(\mu)} $.

\subsubsection{Construction of the phase diagram}

From the previous development it is clear that the nodes in $
\tilde{\Omega}_{max}(\mu) $ may be regarded as ``source'' nodes
which dictate the solution.  To understand how the solution
changes with $ \mu $, we therefore need to determine how $
\tilde{\Omega}_{max}(\mu) $ depends on $ \mu $.

We claim the following:  That there exist a finite number of $ \mu
$, which we denote by $ \mu_1, \dots, \mu_N $, where $ 0 \leq
\mu_1 < \dots < \mu_N < \infty $, for which $ \{(\kappa_{\{i_1,
\dots, i_n\}}, \alpha_{i_1} + \dots + \alpha_{i_n})| \{i_1, \dots,
i_n\} \in \Omega_{max}(\mu)\} $ contains distinct elements. In any
interval $ (\mu_{i-1}, \mu_i) $, $ \Omega_{max}(\mu) $ is
constant, and may therefore be denoted by $ \Omega_i $. The $
\Omega_i $ are all disjoint, and $ \Omega_i \cup \Omega_{i+1}
\subseteq \Omega_{max}(\mu_i) $.

We begin proving this claim by introducing one more definition.
Let $ \Sigma_{\neq} $ denote the set of all sets of nodes, such
that a collection of nodes $ \Omega $ is a member of $
\Sigma_{\neq} $ if and only if $ \{(\kappa_{\{i_1, \dots, i_n\}},
\alpha_{i_1} + \dots + \alpha_{i_n})| \{i_1, \dots, i_n\} \in
\Omega\} $ contains distinct elements.

Note that since there are $ 2^N $ nodes, there are $ 2^{2^N} $
sets of nodes, hence $ \Sigma_{\neq} $ consists of a finite number
of elements.  Given some $ \Omega_{\neq} \in \Sigma_{\neq} $, we
claim that $ \Omega_{max}(\mu) = \Omega_{\neq} $ for at most one $
\mu $.  To show this, suppose that there exist $ \mu_1 < \mu_2 $
for which $ \Omega_{max}(\mu_1) = \Omega_{max}(\mu_2) =
\Omega_{\neq} $.  Choose any two nodes $ \{i_1, \dots, i_n\} $, $
\{j_1, \dots, j_m\} $ in $ \Omega_{\neq} $, and note that $
\kappa_{\{i_1, \dots, i_n\}} e^{-(1 - \alpha_{i_1} - \dots -
\alpha_{i_n}) \mu_1} = \kappa_{\{j_1, \dots, j_m\}} e^{-(1 -
\alpha_{j_1} - \dots - \alpha_{j_m}) \mu_1} = \kappa_{max}(\mu_1)
$, and similarly for $ \mu_2 $.  However, $ a_1 e^{-b_1 x} = a_2
e^{-b_2 x} $ and $ a_1 e^{-b_1 y} = a_2 e^{-b_2 y} $ implies that
$ e^{-b_1 (y-x)} = e^{-b_2 (y-x)} $, so that $ b_1 = b_2 $ and
hence $ a_1 = a_2 $. Therefore, $ \kappa_{\{i_1, \dots, i_n\}} =
\kappa_{\{j_1, \dots, j_m\}} $ and $ \alpha_{i_1} + \dots +
\alpha_{i_n} = \alpha_{j_1} + \dots + \alpha_{j_m} $, so $
\{(\kappa_{\{i_1, \dots, i_n\}}, \alpha_{i_1} + \dots +
\alpha_{i_n})| \{i_1, \dots, i_n\} \in \Omega_{\neq}\} $ does not
contain distinct elements.  Because this contradicts our
assumption about $ \Omega_{\neq} $, it follows that $
\Omega_{max}(\mu) = \Omega_{\neq} $ for at most one $ \mu $.

So, since $ \Sigma_{\neq} $ contains a finite number of elements,
it follows that there are a finite number of $ \mu $ for which $
\Omega_{max}(\mu) $ satisfies the property that $
\{(\kappa_{\{i_1, \dots, i_n\}}, \alpha_{i_1} + \dots +
\alpha_{i_n})| \{i_1, \dots, i_n\} \in \Omega_{max}(\mu)\} $
contains distinct elements.  We can denote these $ \mu $ by $
\mu_1, \dots, \mu_N $, where we assume that $ 0 \leq \mu_1 < \dots
< \mu_N < \infty $.

Note that if a collection of nodes $ \Omega $ has the property
that $ \tilde{\Omega} \neq \Omega $, then $ \Omega $ must be a
collection in $ \Sigma_{\neq} $.  This is easy to see:  $ \Omega $
contains some $ \{i_1, \dots, i_n\} $ for which there exists a
distinct $ \{j_1, \dots, j_m\} \in \Omega $ where $ \{j_1, \dots,
j_m\} \in G_{\{i_1, \dots, i_n\}} $.  Therefore $ \alpha_{i_1} +
\dots + \alpha_{i_n} < \alpha_{j_1} + \dots + \alpha_{j_m} $,
which proves our contention.

We now prove that $ \Omega_{max}(\mu) $ is some constant, which we
denote by $ \Omega_i $, over $ (\mu_{i-1}, \mu_i) $.  Given some $
\mu_0 \in (\mu_{i-1}, \mu_i) $, let $ \mu_+ = \sup\{\tilde{\mu}
\in (\mu_0, \mu_i)| \Omega_{max}(\mu) = \Omega_{max}(\mu_0) \mbox{
} \forall \mbox{ } \mu \in (\mu_0, \tilde{\mu})\} $ ($ \sup $
stands for ``supremum'', which is the least upper bound of a set
of real numbers.  If $ {\bf S} $ is a set of real numbers with an
upper bound, then $ A \equiv \sup {\bf S} $ exists, and satisfies
the following properties:  (1)  $ A $ is an upper bound for $ {\bf
S} $.  (2) If $ B $ is any upper bound of $ {\bf S} $, then $ A
\leq B $. (2) If $ B < A $, then there exists at least one element
of $ {\bf S} $ which exceeds $ B $.). Clearly, $ \mu_+ \leq \mu_i
$. We claim that $ \mu_+ = \mu_i $. To show this, note first of
all that $ \Omega_{max}(\mu) = \Omega_{max}(\mu_0) $ for all $ \mu
\in (\mu_0, \mu_+) $, and that for any $ \tilde{\mu}
> \mu_+ $, there exists $ \mu \in [\mu_+, \tilde{\mu}) $ such that
$ \Omega_{max}(\mu) \neq \Omega_{max}(\mu_0) $.  For given any $
\mu' \in (\mu_0, \mu_+) $, we have, by definition of $ \sup $,
that there exists some $ \tilde{\mu} \in (\mu', \mu_+) $ such that
$ \Omega_{max}(\mu) = \Omega_{max}(\mu_0) $ for all $ \mu \in
(\mu_0, \tilde{\mu}) $. In particular, this implies that $
\Omega_{max}(\mu') = \Omega_{max}(\mu_0) $.  Furthermore, if there
exists $ \tilde{\mu}
> \mu_+ $ for which $ \Omega_{max}(\mu) = \Omega_{max}(\mu_0) $ for
all $ \mu \in [\mu_+, \tilde{\mu}) $, then $ \Omega_{max}(\mu) =
\Omega_{max}(\mu_0) $ for all $ \mu \in (\mu_0, \tilde{\mu}) $,
contradicting the definition of $ \mu_+ $.

Now, suppose $ \Omega_{max}(\mu_+) \notin \Sigma_{\neq} $.  Then
we can write $ \kappa_{\{i_1, \dots, i_n\}} = \kappa_+ $ and $
\alpha_{i_1} + \dots + \alpha_{i_n} = \alpha_+ $ for all $ \{i_1,
\dots, i_n\} \in \Omega_{max}(\mu_+) $.  Then since $
\kappa_{max}(\mu_+) = \kappa_+ e^{-(1 - \alpha_+) \mu_+} $, it
follows by continuity that $ \kappa_+ e^{-(1 - \alpha_+) \mu} >
\kappa_{eff}(\nu; \mu) $ for $ \nu \notin \Omega_{max}(\mu_+) $ in
some neighborhood $ (\mu_+ - \delta, \mu_+ + \delta) $.  But this
implies that $ \Omega_{max}(\mu) = \Omega_{max}(\mu_+) $ over $
(\mu_+ - \delta, \mu_+ + \delta) $.  Since $ \Omega_{max}(\mu_0) =
\Omega_{max}(\mu) $ over $ (\mu_+ - \delta, \mu_+) $, we obtain
that $ \Omega_{max}(\mu) = \Omega_{max}(\mu_0) $ over $ (\mu_0,
\mu_+ + \delta) $, contradicting the definition of $ \mu_+ $.

We have just shown that $ \Omega_{max}(\mu_+) \in \Sigma_{\neq} $.
Since $ \Omega_{max}(\mu) \notin \Sigma_{\neq} $ over $
(\mu_{i-1}, \mu_i) $, we must have that $ \mu_+ = \mu_i $.  Using
a similar argument with $ \inf $, we can show that $
\Omega_{max}(\mu) = \Omega_{max}(\mu_0) $ over $ (\mu_{i-1},
\mu_0) $, and so $ \Omega_{max}(\mu) $ is constant on $
(\mu_{i-1}, \mu_i) $ ($ \inf $ stands for ``infimum'', and is
defined as the greatest lower bound of a set of real numbers.  It
satisfies properties analogous to those of $ \sup $).

Suppose for two $ i, j $ with $ i < j $, we have $ \Omega_i $ and
$ \Omega_j $ are not disjoint.  Then they share at least one node,
and so, by the nature of the two sets, we must have that $
\Omega_i = \Omega_j $.  Define $ \kappa $ to be $ \kappa_{\{i_1,
\dots, i_n\}} $ for any node in $ \Omega_i $, $ \Omega_j $, and $
\alpha $ to be $ \alpha_{i_1} + \dots + \alpha_{i_n} $.  Now, $
\Omega_{max}(\mu_i) $ contains some node $ \{i_1, \dots, i_n\}
\notin \Omega_i $ such that $ \kappa_{eff}(\{i_1, \dots, i_n\};
\mu) < \kappa e^{-(1 - \alpha)\mu} $ for $ \mu $ in $ (\mu_{i-1},
\mu_i) \cup (\mu_{j-1}, \mu_j) $.  But if for $ x_1 < x_2 $ we
have that $ a_1 e^{-b_1 x_1} < a_2 e^{-b_2 x_1} $ and $ a_1
e^{-b_1 x_2} < a_2 e^{-b_2 x_2} $, then $ (a_1/a_2) e^{-(b_1 -
b_2) x_1} < 1 $ and $ (a_1/a_2) e^{-(b_1 - b_2) x_2} < 1 $.  Since
$ (a_1/a_2) e^{-(b_1 - b_2) x} $ is monotone decreasing or
increasing, it follows that $ (a_1/a_2) e^{-(b_1 - b_2) x} < 1 $
on $ (x_1, x_2) $, or equivalently, $ a_1 e^{-b_1 x} < a_2 e^{-b_2
x} $.  Therefore, $ \kappa_{max}(\mu_i) = \kappa_{eff}(\{i_1,
\dots, i_n\}; \mu_i) < \kappa e^{-(1 - \alpha) \mu_i}
\Rightarrow\Leftarrow $.  The $ \Omega_i $ are thus all disjoint,
as claimed.

Finally, since $ \kappa_{max}(\mu) $ is continuous, we have that $
\lim_{\mu \rightarrow \mu_i^-}{\kappa_{max}(\mu)} =
\kappa_{max}(\mu_i) $.  If $ \nu \in \Omega_i $, then this gives $
\kappa_{max}(\mu_i) = \kappa_{eff}(\nu; \mu_i) $.  Similarly,
considering $ \lim_{\mu \rightarrow \mu_i^+}{\kappa_{max}(\mu)} $
gives that $ \kappa_{max}(\mu_i) = \kappa_{eff}(\nu; \mu_i) $ for
$ \nu \in \Omega_{i+1} $.  Therefore, $ \Omega_i, \Omega_{i+1}
\subseteq \Omega_{max}(\mu_i) $, so $ \Omega_i \cup \Omega_{i+1}
\subseteq \Omega_{max}(\mu_i) $, as claimed.

The various $ \Omega_i $ may therefore be regarded as defining
different ``phases'' in the equilibrium behavior of the model.
Physically, each ``phase'' is defined by a set of ``source
nodes,'' which dictate which genes in the genome are knocked out,
and which are not.  The transition from $ \Omega_i $ to $
\Omega_{i+1} $ corresponds to certain genes in the genome becoming
knocked out, and perhaps other genes becoming viable again. This
transition can happen more than once, and so we refer to the
series of $ \Omega_i \rightarrow \Omega_{i+1} $ transitions as an
``error cascade.''

Because $ \kappa_{eff}(\{1, \dots, N\}; \mu) = 1 $, for
sufficiently large $ \mu $, $ \kappa_{eff}(\{1, \dots, N\}; \mu) >
\kappa_{eff}(\nu; \mu) $ for any $ \nu \neq \{1, \dots, N\} $.
Therefore, for sufficiently large $ \mu $, $ \Omega_{max}(\mu) =
\{\{1, \dots, N\}\} $.  Since $ \Omega_{max}(\mu) $ is constant on
$ (\mu_N, \infty) $, it follows that $ \Omega_{max}(\mu) = \{\{1,
\dots, N\}\} $ on $ (\mu_N, \infty) $.  Thus, the final transition
from $ \Omega_{N} $ to $ \Omega_{N+1} $ corresponds to
delocalization over the entire genome space, which is simply the
error catastrophe.

\subsubsection{Finding the $ z_{l_1, \dots, l_N} $}

Once we have determined $ \bar{\kappa}(t = \infty) $, we can in
principle obtain the population fractions $ z_{l_1, \dots, l_N} $
in the various Hamming classes.  The problem is that, if $
z_\emptyset = 0 $, then for any {\it finite} values of $ l_1,
\dots, l_n $, we get that $ z_{l_1, \dots, l_N} = 0 $.  To show
this, suppose we can find $ l_1, \dots, l_N $ such that $ z_{l_1,
\dots, l_N} > 0 $ at equilibrium.  Of the $ l_1, \dots, l_N $ for
which $ z_{l_1, \dots, l_N} > 0 $, choose a set of indices $ l_1',
\dots, l_N' $ such that $ l_1' + \dots + l_N' $ is as small as
possible. Note that if $ z_{l_1, \dots, l_N} = z_{l_1' - l_1'',
\dots, l_N' - l_N''} $, with $ (l_1'', \dots, l_N'') \neq (0,
\dots, 0) $, then $ z_{l_1, \dots, l_N} = 0 $.

Now, let the nonzero $ l_i' $ be denoted by $ l_{i_1}', \dots,
l_{i_n}' $. Then $ \kappa_{l_1', \dots, l_N'} = \kappa_{\{i_1,
\dots, i_n\}} $, and we have, from Eq. (11), that, at equilibrium,
\begin{equation}
0 = \frac{d z_{l_1', \dots, l_N'}}{dt}\lvb_{t = \infty} =
(\kappa_{\{i_1, \dots, i_n\}} e^{-\mu} - \bar{\kappa}(t = \infty))
z_{l_1', \dots, l_N'}
\end{equation}
which gives $ \bar{\kappa}(t = \infty) = \kappa_{\{i_1, \dots,
i_n\}} e^{-\mu} $.  But, $ \bar{\kappa}(t = \infty) \geq
\kappa_{\{i_1, \dots, i_n\}} e^{-(1 - \alpha_{i_1} - \dots -
\alpha_{i_n})\mu} $.  Therefore, $ e^{-\mu} \geq e^{-(1 -
\alpha_{i_1} - \dots - \alpha_{i_n}) \mu} $, and so $ \alpha_{i_1}
+ \dots + \alpha_{i_n} = 0 $, hence $ n = 0 $.  But then $
z_{l_1', \dots, l_N'} = z_\emptyset > 0 \Rightarrow\Leftarrow $.
This proves our claim.

If $ \tilde{\Omega}_{max}(\mu) = \emptyset $, then the above claim
does not present us with any problem.  We can simply recursively
solve Eq. (11) at equilibrium for all the $ z_{l_1, \dots, l_N} $.
But once any delocalization occurs, it is impossible to solve for
the equilibrium distribution in terms of the Hamming classes.
However, we can recursively obtain the distribution of another
class of population fractions, as follows:  Given some collection
of indices $ \{i_1, \dots, i_n\} $, another collection of indices
$ \{j_1, \dots, j_k\} \subseteq \{i_1, \dots, i_n\} $, and a
collection of Hamming distances $ l_1, \dots, l_N $, we define $
\tilde{z}_{\{j_1, \dots, j_k\}}(\vec{l}_{\{i_1, \dots, i_n\}}) $
and $ z_{\{j_1, \dots, j_k\}}(\vec{l}_{\{i_1, \dots, i_n\}}) $ as,
\begin{widetext}
\begin{eqnarray}
&   & \tilde{z}_{\{j_1, \dots, i_k\}}(\vec{l}_{\{i_1, \dots,
i_n\}}) = \sum_{l_{j_1} = 1}^{\infty} \cdot \dots \cdot
\sum_{l_{j_k} = 1}^{\infty}{z_{l_{j_1} {\bf e}_{j_1} + \dots +
l_{j_k} {\bf e}_{j_k} + \sum_{i \in \{1, \dots, N\}/\{i_1, \dots,
i_n\}} l_i {\bf e}_i}}
\nonumber \\
&   & z_{\{j_1, \dots, j_k\}}(\vec{l}_{\{i_1, \dots, i_n\}}) =
\sum_{l_{j_1} = 0}^{\infty} \cdot \dots \cdot \sum_{l_{j_k} =
0}^{\infty}{z_{l_{j_1} {\bf e}_{j_1} + \dots + l_{j_k} {\bf
e}_{j_k} + \sum_{i \in \{1, \dots, N\}/\{i_1, \dots, i_n\}} l_i
{\bf e}_i}}
\end{eqnarray}
It is possible to show that,
\begin{eqnarray}
z_{\{j_1, \dots, j_k\}}(\vec{l}_{\{i_1, \dots, i_n\}}) & = &
\sum_{l = 0}^{k} \sum_{\{j_1', \dots, j_l'\} \subseteq \{j_1,
\dots, j_k\}} \tilde{z}_{\{j_1', \dots, j_l'\}}(\vec{l}_{\{i_1,
\dots, i_n\}})
\end{eqnarray}
and hence that,
\begin{eqnarray}
\tilde{z}_{\{j_1, \dots, j_k\}}(\vec{l}_{\{i_1, \dots, i_n\}}) & =
& \sum_{l = 0}^{k} (-1)^{k-l} \sum_{\{j_1', \dots, j_l'\}
\subseteq \{j_1, \dots, j_k\}} z_{\{j_1', \dots,
j_l'\}}(\vec{l}_{\{i_1, \dots, i_n\}})
\end{eqnarray}
We may then derive the expression for $ d \tilde{z}_{\{i_1, \dots,
i_n\}}(\vec{l}_{\{i_1, \dots, i_n\}})/dt $.  Since the derivation
uses techniques similar to those used in Appendices A and B, we
simply state the final result, which is,
\begin{eqnarray}
\frac{d \tilde{z}_{\{i_1, \dots, i_n\}}(\vec{l}_{\{i_1, \dots,
i_n\}})}{dt} & = & e^{-(1 - \alpha_{i_1} - \dots -
\alpha_{i_n})\mu} \sum_{k = 0}^{n} \sum_{\{j_1, \dots, j_k\}
\subseteq \{i_1, \dots, i_n\}} \sum_{\stackrel{l_i' = 0}{i \in
\{1, \dots, N\}/\{i_1, \dots, i_n\}}}^{l_i} \prod_{i \in \{1,
\dots, N\}/\{i_1, \dots,
i_n\}}{\frac{(\alpha_i \mu)^{l_i'}}{l_i'!}} \times \nonumber \\
&   & \Pi_{j \in \{i_1, \dots, i_n\}/\{j_1, \dots, j_k\}} (1 -
e^{-\alpha_j \mu}) \times \nonumber \\
&   & \kappa_{\{j_1, \dots, j_k\}}(\vec{l}_{\{i_1, \dots, i_n\}} -
\vec{l}'_{\{i_1, \dots, i_n\}}) \tilde{z}_{\{j_1, \dots,
j_k\}}(\vec{l}_{\{i_1, \dots, i_n\}} - \vec{l}'_{\{i_1, \dots,
i_n\}}) \nonumber \\
&   & - \bar{\kappa}(t) \tilde{z}_{\{i_1, \dots,
i_n\}}(\vec{l}_{\{i_1, \dots, i_n\}})
\end{eqnarray}
\end{widetext}
where $ \kappa_{\{j_1, \dots, j_k\}}(\vec{l}_{\{i_1, \dots,
i_n\}}) = \kappa_{\{j_1, \dots, j_k\} \cup \{j_1', \dots, j_l'\}}
$, where $ \{j_1', \dots, j_l'\} $ are the indices of the nonzero
Hamming distances in $ \vec{l}_{\{i_1, \dots, i_n\}} $.

We claim that, at equilibrium, $ \tilde{z}_{\nu}(\vec{l}_\nu)
> 0 $ only if $ \nu \in G_{\tilde{\nu}} $ for some
$ \tilde{\nu} \in \tilde{\Omega}_{max}(\mu) $ for which $
\tilde{z}_{\tilde{\nu}} > 0 $. For if $
\tilde{z}_{\nu}(\vec{l}_\nu) > 0 $, let $ \tilde{\nu} = \{i_1,
\dots, i_n\} \subseteq \nu $ be a node of minimal level for which
there exists $ \vec{l}_{\tilde{\nu}} $ such that $
\tilde{z}_{\tilde{\nu}}(\vec{l}_{\tilde{\nu}}) > 0 $. Note then
that for any proper subset $ \{j_1, \dots, j_k\} \subset \{i_1,
\dots, i_n\} $, we must have that $ \tilde{z}_{\{j_1, \dots,
j_k\}}(\vec{l}_{\{i_1, \dots, i_n\}}) = 0 $.  This implies that,
at equilibrium,
\begin{eqnarray}
0 & = & \frac{d \tilde{z}_{\{i_1, \dots, i_n\}}(\vec{l}_{\{i_1,
\dots, i_n\}})}{dt} \nonumber \\
& = & e^{-(1 - \alpha_{i_1} - \dots -
\alpha_{i_n})\mu} \times \nonumber \\
&   & \sum_{\stackrel{l_i' = 0}{i \in \{1, \dots, N\}/\{i_1,
\dots, i_n\}}}^{l_i} \prod_{i \in \{1, \dots, N\}/\{i_1, \dots,
i_n\}}{\frac{(\alpha_i \mu)^{l_i'}}{l_i'!}} \times
\nonumber \\
&   & \kappa_{\{i_1, \dots, i_n\}}(\vec{l}_{\{i_1, \dots, i_n\}} -
\vec{l}'_{\{i_1, \dots, i_n\}}) \times \nonumber
\\
&   & \tilde{z}_{\{i_1, \dots, i_n\}}(\vec{l}_{\{i_1,
\dots, i_n\}} - \vec{l}'_{\{i_1, \dots, i_n\}}) \nonumber \\
&   & - \bar{\kappa}(t = \infty) \tilde{z}_{\{i_1, \dots,
i_n\}}(\vec{l}_{\{i_1, \dots, i_n\}})
\end{eqnarray}
Among all $ \vec{l}_{\{i_1, \dots, i_n\}} $ for which $
\tilde{z}_{\{i_1, \dots, i_n\}}(\vec{l}_{\{i_1, \dots, i_n\}}) > 0
$, there exists an $ \vec{l}''_{\{i_1, \dots, i_n\}} $ such that $
\sum_{i \in \{1, \dots, N\}/\{i_1, \dots, i_n\}}{l_i''} $ is
minimal.  Then we obtain,
\begin{eqnarray}
0 & = & \frac{d \tilde{z}_{\{i_1, \dots, i_n\}}(\vec{l}''_{\{i_1,
\dots, i_n\}})}{dt}\lvb_{t = \infty} \nonumber \\
& = & (\kappa_{\{i_1, \dots, i_n\}}(\vec{l}''_{\{i_1, \dots,
i_n\}}) e^{-(1 - \alpha_{i_1} - \dots - \alpha_{i_n})\mu} -
\bar{\kappa}(t)) \times \nonumber \\
&   & \tilde{z}_{\{i_1, \dots, i_n\}}(\vec{l}''_{\{i_1, \dots,
i_n\}})
\end{eqnarray}
which gives $ \bar{\kappa}(t = \infty) = \kappa_{\{i_1, \dots,
i_n\}}(\vec{l}''_{\{i_1, \dots, i_n\}}) e^{-(1 - \alpha_{i_1} -
\dots - \alpha_{i_n})\mu} $.  Now, let $ i_1', \dots, i_m' $
denote the indices of the nonzero Hamming distances in $
\vec{l}_{\{i_1, \dots, i_n\}} $.  Then $ \kappa_{\{i_1, \dots,
i_n\}} = \kappa_{\{i_1, \dots, i_n\} \cup \{i_1', \dots, i_m'\}}
$.  But since $ \bar{\kappa}(t = \infty) \geq \kappa_{\{i_1,
\dots, i_n\} \cup \{i_1', \dots, i_m'\}} e^{-(1 - \alpha_{i_1} -
\dots - \alpha_{i_n} - \alpha_{i_1'} - \dots - \alpha_{i_m'})\mu}
$, we get $ \alpha_{i_1'} + \dots + \alpha_{i_m'} = 0 $, so $ m =
0 $.  Therefore $ \bar{\kappa}(t = \infty) =
\kappa_{eff}(\tilde{\nu}; \mu) $, so since $
\tilde{z}_{\tilde{\nu}} > 0 $, we have $ \tilde{\nu} \in
\tilde{\Omega}_{max}(\mu) $.

The $ \tilde{z}_{\nu}(\vec{l}_{\nu}) $ may be obtained recursively
from Eq. (27), starting with the values of $ \tilde{z}_\nu $ for $
\nu \in \tilde{\Omega}_{max}(\mu) $.  The idea is that, starting
with the values of $ \tilde{z}_\nu $ for $ \nu \in
\tilde{\Omega}_{max}(\mu) $, we may compute $
\tilde{z}_\nu(\vec{l}_\nu) $ recursively.  We then proceed down
the levels, computing the $ \tilde{z}_\nu(\vec{l}_\nu) $ using the
values of $ \tilde{z}_\nu(\vec{l}_\nu - \vec{l}'_\nu) $ and $
\tilde{z}_{\tilde{\nu}}(\vec{l}_{\tilde{\nu}}) $ for $ \tilde{\nu}
\subset \nu $.  Note then that instead of computing the $ z_{l_1,
\dots, l_N} $, which will be $ 0 $ as soon as any delocalization
occurs, we first sum over a set of gene indices containing the
delocalized genes as a subset, and only compute the population
distribution for finite Hamming distances of the localized genes.

\subsection{Localization Lengths}

In this subsection, we compute various localization lengths
associated with the population distribution.  Specifically, given
a node $ \{i_1, \dots, i_n\} $, and some $ i \notin \{i_1, \dots,
i_n\} $, we define two localization lengths, $ \langle
l_i\rangle_{\{i_1, \dots, i_n\}} $ and $ \tilde{\langle l_i
\rangle}_{\{i_1, \dots, i_n\}} $, as follows:
\begin{eqnarray}
&   & \langle l_i\rangle_{\{i_1, \dots, i_n\}} \equiv
\sum_{l_{i_1} = 0}^{\infty} \cdot \dots \cdot \sum_{l_{i_n} =
0}^{\infty} \sum_{l_i = 1}^{\infty} {l_i z_{l_{i_1} {\bf e}_{i_1}
+ \dots + l_{i_n} {\bf e}_{i_n} + l_i {\bf e}_i}}
\nonumber \\
\\
&   & \tilde{\langle l_i\rangle}_{\{i_1, \dots, i_n\}} \equiv
\sum_{l_{i_1} = 1}^{\infty} \cdot \dots \cdot \sum_{l_{i_n} =
1}^{\infty} \sum_{l_i = 1}^{\infty} {l_i z_{l_{i_1} {\bf e}_{i_1}
+ \dots + l_{i_n} {\bf e}_{i_n} + l_i {\bf e}_i}} \nonumber \\
\end{eqnarray}
Note that,
\begin{equation}
\langle l_i\rangle_{\{i_1, \dots, i_n\}} =
\sum_{k = 0}^{n}\sum_{\{j_1, \dots, j_k\} \subseteq \{i_1, \dots, i_n\}}
{\tilde{\langle l_i\rangle}_{\{j_1, \dots, j_k\}}}
\end{equation}
and so, in analogy with $ z_{\{i_1, \dots, i_n\}} $ and
$ \tilde{z}_{\{i_1, \dots, i_n\}} $, we have that,
\begin{equation}
\tilde{\langle l_i\rangle}_{\{i_1, \dots, i_n\}} = \sum_{k =
0}^{n} (-1)^{n-k} \sum_{\{j_1, \dots, j_k\} \subseteq \{i_1,
\dots, i_n\}} {\langle l_i \rangle_{\{j_1, \dots, j_k\}}}
\end{equation}
We also define the localization length $ \langle l_i\rangle $ by,
\begin{equation}
\langle l_i \rangle =
\sum_{l_1 = 0}^{\infty} \cdot \dots \cdot
\sum_{l_N = 0}^{\infty}{l_i z_{l_1, \dots, l_N}}
\end{equation}
Note that $ \langle l_i \rangle = \langle l_i \rangle_{\{1, \dots,
N\}/\{i\}} = \sum_{n = 0}^{N-1} \sum_{\{i_1, \dots, i_n\}
\subseteq \{1, \dots, N\}/\{i\}} \tilde{\langle
l_i\rangle}_{\{i_1, \dots, i_n\}} $, and so is finite if and only
if all the $ \tilde{\langle l_i\rangle}_{\{i_1, \dots, i_n\}} $
are finite.

We can compute $ \tilde{\langle l_i \rangle}_{\{i_1, \dots, i_n\}} $ at
equilibrium by finding the time derivative and setting it to $ 0 $.  In
Appendix B we show that,
\begin{widetext}
\begin{eqnarray}
\frac{d \tilde{\langle l_i\rangle}_{\{i_1, \dots, i_n\}}}{dt}
& = &
(\kappa_{eff}(\{i_1, \dots, i_n, i\}; \mu) - \bar{\kappa}(t))
\tilde{\langle l_i\rangle}_{\{i_1, \dots, i_n\}} \nonumber \\
&   &
+ \alpha_i \mu e^{-(1 - \alpha_{i_1} - \dots - \alpha_{i_n} - \alpha_i)\mu}
(\kappa_{\{i_1, \dots, i_n\}} \tilde{z}_{\{i_1, \dots, i_n\}} +
 \kappa_{\{i_1, \dots, i_n, i\}} \tilde{z}_{\{i_1, \dots, i_n, i\}})
\nonumber \\
&   & + e^{-(1 - \alpha_{i_1} - \dots - \alpha_{i_n} -
\alpha_i)\mu} \sum_{k = 0}^{n-1} \sum_{\{j_1, \dots, j_k\} \subset
\{i_1, \dots, i_n\}}
\times \nonumber \\
&   & (\kappa_{\{j_1, \dots, j_k, i\}} \tilde{\langle l_i
\rangle}_{\{j_1, \dots, j_k\}} +
 \alpha_i \mu
 \kappa_{\{j_1, \dots, j_k\}} \tilde{z}_{\{j_1, \dots, j_k\}} +
 \alpha_i \mu
 \kappa_{\{j_1, \dots, j_k, i\}} \tilde{z}_{\{j_1, \dots, j_k, i\}})
\times \nonumber \\
&   & \prod_{j \in \{i_1, \dots, i_n\}/\{j_1, \dots, j_k\}} (1 -
e^{-\alpha_j \mu})
\end{eqnarray}
Therefore, at equilibrium, we get,
\begin{eqnarray}
\tilde{\langle l_i\rangle}_{\{i_1, \dots, i_n\}}
& = &
\alpha_i \mu
\frac{e^{-(1 - \alpha_{i_1} - \dots - \alpha_{i_n} - \alpha_i) \mu}}
     {\bar{\kappa}(t = \infty) - \kappa_{eff}(\{i_1, \dots, i_n, i\}; \mu)}
(\kappa_{\{i_1, \dots, i_n\}} \tilde{z}_{\{i_1, \dots, i_n\}} +
 \kappa_{\{i_1, \dots, i_n, i\}} \tilde{z}_{\{i_1, \dots, i_n, i\}})
\nonumber \\
&   &
+ \frac{e^{-(1 - \alpha_{i_1} - \dots - \alpha_{i_n} - \alpha_i)\mu}}
       {\bar{\kappa}(t = \infty) - \kappa_{eff}(\{i_1, \dots, i_n, i\}; \mu)}
\sum_{k = 0}^{n-1}\sum_{\{j_1, \dots, j_k\} \subset \{i_1, \dots,
i_n\}}
\times \nonumber \\
&   & (\kappa_{\{j_1, \dots, j_k, i\}} \tilde{\langle l_i
\rangle}_{\{j_1, \dots, j_k\}} +
 \alpha_i \mu
 \kappa_{\{j_1, \dots, j_k\}} \tilde{z}_{\{j_1, \dots, j_k\}} +
 \alpha_i \mu
 \kappa_{\{j_1, \dots, j_k, i\}} \tilde{z}_{\{j_1, \dots, j_k, i\}})
\times \nonumber \\
&   & \prod_{j \in \{i_1, \dots, i_n\}/\{j_1, \dots, j_k\}} (1 -
e^{-\alpha_j \mu})
\end{eqnarray}
\end{widetext}

We can characterize the behavior of the $ \tilde{\langle
l_i\rangle}_{\{i_1, \dots, i_n\}} $.  First of all, we claim that
$ \tilde{\langle l_i\rangle}_{\{i_1, \dots, i_n\}} = 0 $ if and
only if $ \tilde{z}_{\{i_1, \dots, i_n, i\}} = 0 $.  Secondly, we
claim that $ \tilde{\langle l_i\rangle}_{\{i_1, \dots, i_n\}} =
\infty $ if and only if $ \{j_1, \dots, j_k, i\} \in
\tilde{\Omega}_{max}(\mu) $ with $ \tilde{z}_{\{j_1, \dots, j_k,
i\}} > 0 $ for some $ \{j_1, \dots, j_k\} \subseteq \{i_1, \dots,
i_n\} $.

To show this, note first of all that, from physical considerations,
$ \tilde{\langle l_i\rangle}_{\{i_1, \dots, i_n\}} = 0 $ if
$ \tilde{z}_{\{i_1, \dots, i_n, i\}} = 0 $.  If
$ \tilde{z}_{\{i_1, \dots, i_n, i\}} > 0 $, then
$ \{i_1, \dots, i_n, i\} \in G_{\tilde{\Omega}_{max}(\mu)} $,
and so, since $ \bar{\kappa}(t = \infty) \geq
\kappa_{eff}(\{i_1, \dots, i_n\}; \mu) $, it follows that
$ \tilde{\langle l_i \rangle}_{\{i_1, \dots, i_n\}} > 0 $.  This
establishes the first part of our claim.

So now suppose that $ \{j_1, \dots, j_k, i\} \in \tilde{\Omega}_{max}(\mu) $,
with $ \tilde{z}_{\{j_1, \dots, j_k, i\}} > 0 $ for some $ \{j_1, \dots, j_k\}
\subseteq \{i_1, \dots, i_n\} $.  Then $ \bar{\kappa}(t = \infty) =
\kappa_{eff}(\{j_1, \dots, j_k, i\}; \mu) $, and so,
\begin{eqnarray}
\tilde{\langle l_i\rangle}_{\{j_1, \dots, j_k\}} & = & \alpha_i
\mu \frac{e^{-(1 - \alpha_{j_1} - \dots - \alpha_{j_k} -
\alpha_i)\mu}} {\bar{\kappa}(t = \infty) - \kappa_{eff}(\{j_1,
\dots, j_k, i\}; \mu)} \times \nonumber \\
&   &
\kappa_{\{j_1, \dots, j_k, i\}} \tilde{z}_{\{j_1, \dots, j_k,
i\}} = \infty
\end{eqnarray}
which of course implies that $ \tilde{\langle l_i\rangle}_{\{i_1, \dots, i_n\}}
= \infty $.

To prove the converse, let us suppose that $ \tilde{\langle
l_i\rangle}_{\{i_1, \dots, i_n\}} = \infty $. Let us choose $
\{j_1, \dots, j_k\} \subseteq \{i_1, \dots, i_n\} $ to be the
minimal level subset for which $ \tilde{\langle
l_i\rangle}_{\{j_1, \dots, j_k\}} = \infty $. Then if $
\bar{\kappa}(t = \infty) > \kappa_{eff}(\{j_1, \dots, j_k, i\};
\mu) $, it is clear from the expression for $ d \tilde{\langle
l_i\rangle}_ {\{j_1, \dots, j_k\}}/dt $ that $ \tilde{\langle
l_i\rangle}_ {\{j_1', \dots, j_l'\}} = \infty $ for some $ \{j_1',
\dots, j_l'\} \subset \{j_1, \dots, j_k\} $, with $ 0 \leq l \leq
k-1 $. But this contradicts the minimality of $ k $, hence $
\bar{\kappa}(t = \infty) = \kappa_{eff}(\{j_1, \dots, j_k, i\};
\mu) $, so since $ \tilde{z}_{\{j_1, \dots, j_k, i\}} > 0 $, it
follows that $ \{j_1, \dots, j_k, i\} \in
\tilde{\Omega}_{max}(\mu) $. This proves the converse, which
establishes the second part of our claim.

\subsection{A Simple Example}

We now illustrate the theory developed above using a simple
two-gene ``network'' as an example.  We assume a genome containing
two identical genes, so that $ \alpha_1 = \alpha_2 = 1/2 $, and we
choose the following growth parameters:  $ \kappa_\emptyset = 10
$, $ \kappa_{\{1\}} = \kappa_{\{2\}} = 5 $, and $ \kappa_{\{1,2\}}
= 1 $.

With these parameters, the system exhibits two localization to
delocalization transitions.  First, for $ \mu \in [0, 2 \ln 2) $
we have $ \tilde{\Omega}_{max}(\mu) = \emptyset $.  For $ \mu \in
(2 \ln 2, 2 \ln 5) $ we have $ \tilde{\Omega}_{max}(\mu) =
\{\{1\}, \{2\}\} $. The error catastrophe occurs at $ \mu = 2 \ln
5 $.

We determined the equilibrium behavior of the model by solving the
finite sequence length equations for $ L = 40 $ and $ S = 2 $. The
details may be found in Appendix C.  Figure 4 shows a plot of $
\bar{\kappa}(t = \infty) $ from the simulation results and from
our theory.  Figure 5 shows plots of $ \tilde{z}_\emptyset $, $
\tilde{z}_{\{1\}} $, $ \tilde{z}_{\{2\}} $, and $
\tilde{z}_{\{1,2\}} $ from the simulation results and from theory.

\begin{figure}
\includegraphics[width = 0.9\linewidth]{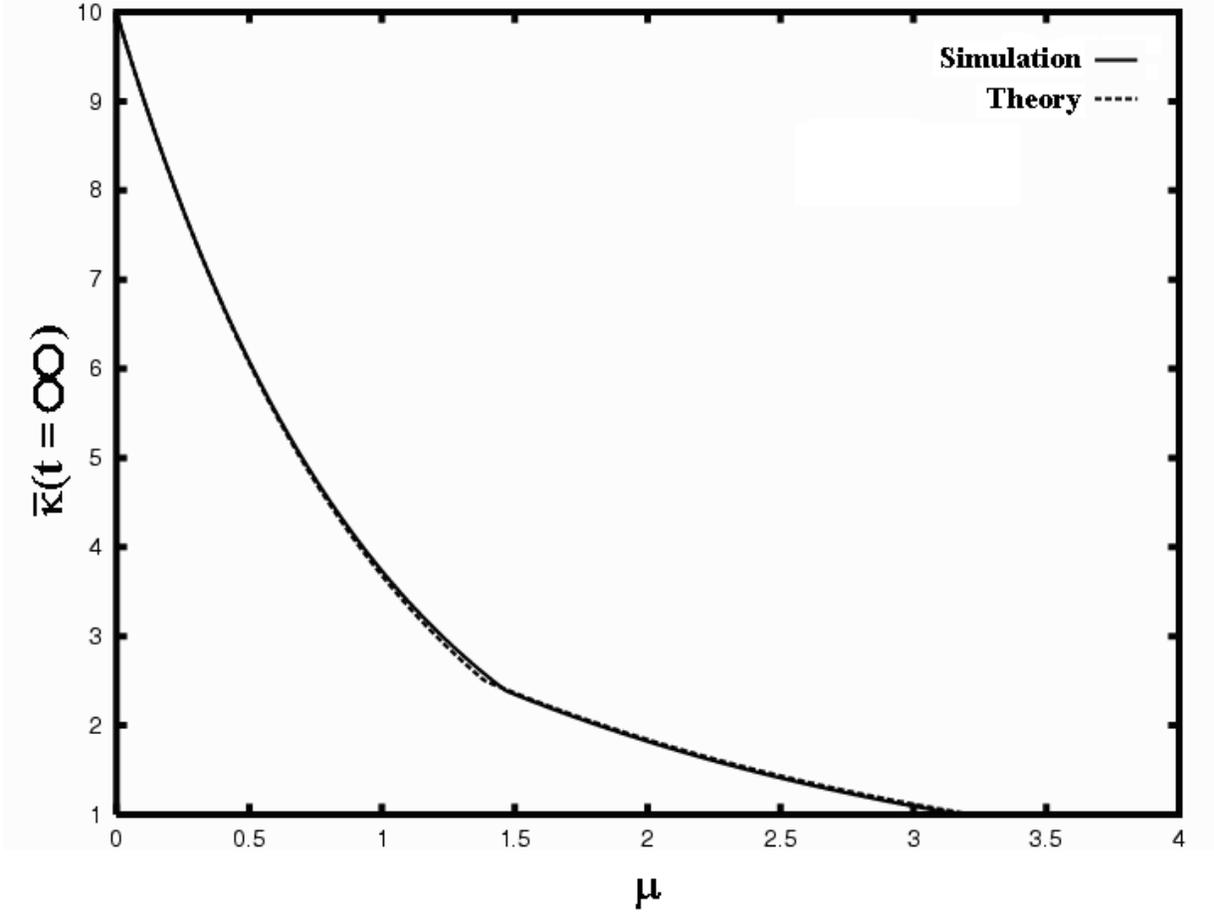}
\caption{Plot of $ \bar{\kappa}(t = \infty) $ from both simulation
and theory.}
\end{figure}

\begin{figure}
\includegraphics[width = 0.9\linewidth]{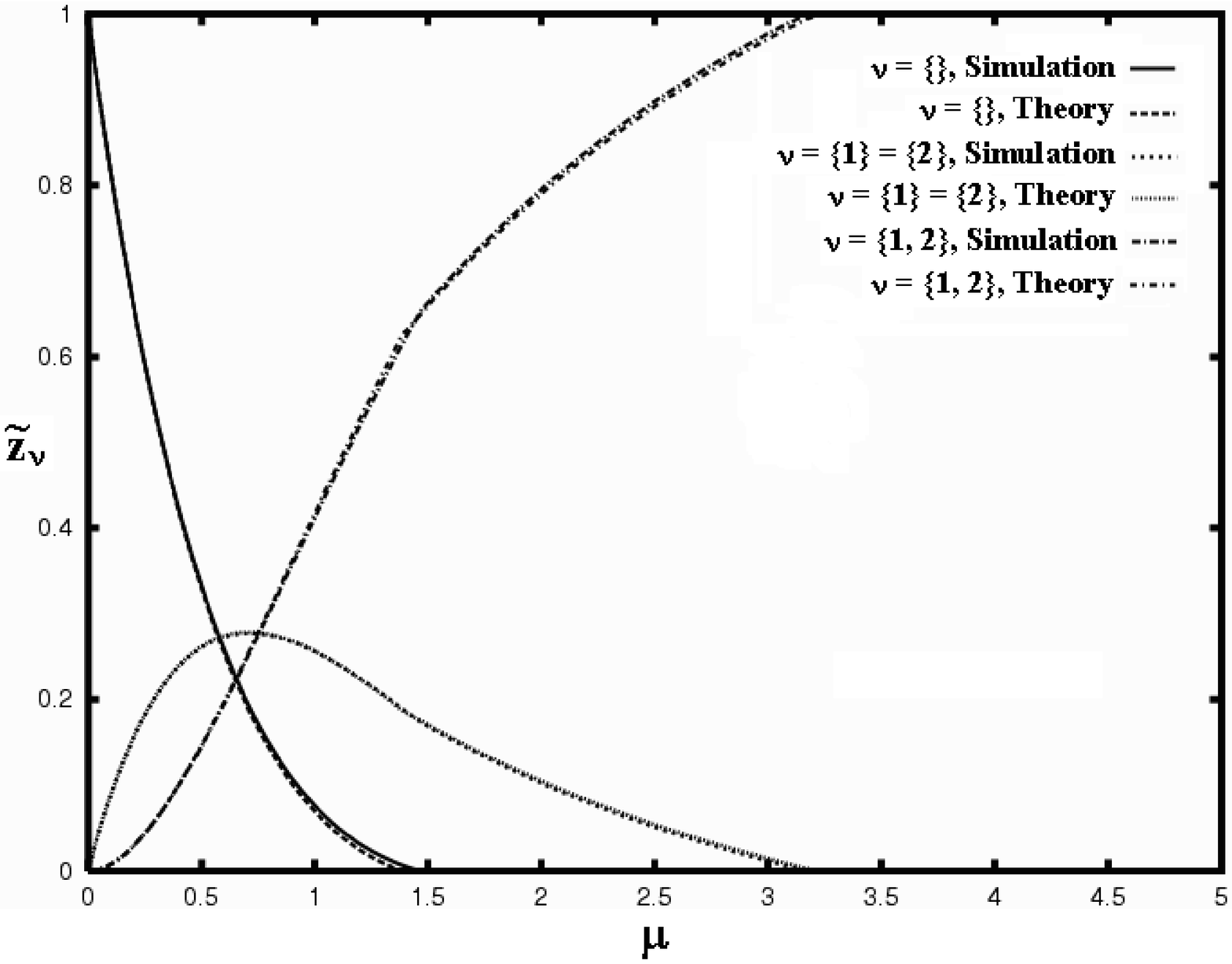}
\caption{Plots of $ \tilde{z}_\emptyset $, $ \tilde{z}_{\{1\}} $,
$ \tilde{z}_{\{2\}} $, and $ \tilde{z}_{\{1, 2\}} $ from both
simulation and theory.  By symmetry, $ w_{\{1\}} = w_{\{2\}} = 1/2
$ when $ \tilde{\Omega}_{max}(\mu) = \{\{1\}, \{2\}\} $.}
\end{figure}

\section{Discussion}

The first point to note about the solution of the quasispecies
equations for a gene network is that, unlike the single gene
model, which exhibits a single ``error catastrophe,'' the multiple
gene model exhibits a series of localization to delocalization
transitions which we term an ``error cascade.'' The reason for
this is that as the mutation rate is increased, the selective
advantage for maintaining functional copies of certain genes in
the genome is no longer sufficiently strong to localize the
population distribution about the corresponding master sequences,
and delocalization occurs in the corresponding sequence spaces.

The more a given gene or set of genes contributes to the fitness
of an organism, the larger $ \mu $ will have to be to induce
delocalization in the corresponding sequence spaces.  Eventually,
by making $ \mu $ sufficiently large, the selective advantage for
maintaining any functional genes in the genome will disappear, and
the result is complete delocalization over all sequence spaces,
corresponding to the error catastrophe.

The prediction of an error cascade suggests an approach for
determining the selective advantage of maintaining certain genes
in a genome.  Currently, the standard method for determining
whether a gene is ``essential'' is by knocking it out, and then
seeing if the organism survives.  By knocking out each of the
genes, one can construct a ``deletion set'' for a given organism,
consisting of the minimal set of genes necessary for the organism
to survive \cite{DELSET}.

While knowledge of the deletion set of an organism is important,
it does not explain why the organism should maintain functional
copies of other, ``nonessential'' genes.  One possibility is that
these ``nonessential'' do confer a fitness advantage to the
organism, however, the time scale on which organisms are observed
to grow during knockout experiments is simply too short to resolve
these fitness differences.

Thus, an alternative approach to the deletion set method is to
have organisms grow at various mutagen concentrations.  By determining
which genes get knocked out at the corresponding mutation rates,
it is possible to determine the relative importance of various
genes to the fitness of an organism.  Such an experiment is
likely to be difficult to perform.  Nevertheless, if successful,
it would provide a considerably more powerful approach than
the deletion set method for determining fitness advantages of
various genes.

The results in this paper also shed light on a phenomenon which
C.O. Wilke termed the ``survival of the flattest'' \cite{WILKE1}.
Briefly, what Wilke (and others) showed was that at low mutation
rates, a population will localize in a region of sequence space
which has high fitness.  At higher mutation rates, a population
will relocalize in a region of sequence space which may not have
maximal fitness, but is mutationally robust \cite{WILKE1}.

The error cascade is exactly a relocalization from a region of
high fitness but low mutational support to a region of lower
fitness but higher mutational support.  The reason for this is
that the fitness landscape becomes progressively flatter as more
and more genes are knocked out, because the more genes are knocked
out, the smaller the fraction of the genome which is involved in
determining the fitness of the organism.

This implies that an error cascade is necessary for the ``survival
of the flattest'' principle to hold.  Robustness in this sense is
therefore conferred by modularity in the genome.  That is,
robustness does not arise because an individual gene may remain
functional after several point-mutations, but rather arises from
the fact that the organism can remain viable even if entire
regions (e.g. ``genes'') of the genome are knocked out (it should
be noted that the idea that mutational robustness is conferred by
modularity in the genome has been discussed before \cite{WILKE1}).

To see this point more clearly, one can consider a ``robust''
landscape in which the genome consists of a single gene.  However,
unlike the single-fitness peak landscape, the organism is viable
out to some Hamming class $ l_{via} $.  Therefore, if $
D_H(\sigma, \sigma_0) = l $, then $ \kappa_{\sigma} = 1 $ if $ l >
l_{via} $, otherwise $ \kappa_{\sigma} = \kappa_l $, where $
\kappa_0 \geq \kappa_1 \geq \dots \geq \kappa_{l_{via}} > 1 $.
Using techniques similar to the ones used in this paper (neglect
of backmutations and stability criterion for equilibria), it is
possible to show that the equilibrium mean fitness is exactly $
\kappa_0 e^{-\mu} $, unchanged from the single-fitness peak
results.  Thus, in contrast to robustness studies which consider
finite sequence lengths and do not have a well-defined viability
cutoff \cite{ROBUSTPNAS}, in the limit of infinite sequence length
there is no selective advantage in having a genome which can
sustain a finite number of point mutations and remain viable.

\section{Conclusions}

This paper developed an extension of the quasispecies model for
arbitrary gene networks.  We considered the case of conservative
replication and a genome-independent replication error
probability.  We showed that, instead of a single error
catastrophe, the model exhibits a series of localization to
delocalization transitions, termed an ``error cascade.''

While the numerical example we used in this paper was relatively
simple (in order to clearly illustrate the theory developed), it
is possible to have nontrivial delocalization behavior, depending
on the choice of the landscape.  For example, it is possible that
certain genes which are knocked out in one phase can become
reactivated again in the following phase.  That is, instead of a
delocalization, one can have a {\it re-localization} to source
nodes not contained in the mutational subgraphs of the source
nodes in the previous phase.  This implies that the $
\tilde{z}_{\nu} $ can exhibit discontinuous behavior.  The types
of equilibrium behaviors possible is something which will be
explored in future research.

Future research also will involve incorporating more details to
the multiple-gene model introduced in this paper.  For example,
one extension is to move away from the ``single-fitness peak''
assumption for each gene.  Another natural extension is to study
the equilibrium behavior of the multiple-gene quasispecies
equations for the case of semiconservative replication.  While
this is a subject for future work, we hypothesize that many of the
semiconservative results would be essentially unchanged from the
conservative ones.  Thus, we claim that at equilibrium, we would
still have that $ \bar{\kappa}(t = \infty) = \kappa_{max}(\mu) $,
only this time $ \kappa_{eff}(\nu; \mu) $ is computed by replacing
$ e^{-\mu} $ with $ 2 e^{-\mu/2} - 1 $.  We also claim that we
would still have that $ \tilde{\Omega}_{max}(\mu) $ define the
``source'' nodes of the equilibrium solution.  Indeed, we
hypothesize that, for semiconservative replication, Eq. (13)
becomes,
\begin{eqnarray}
\frac{d \tilde{z}_{\{i_1, \dots, i_n\}}}{dt} & = &
(\kappa_{eff}(\{i_1, \dots, i_n\}; \mu) - \bar{\kappa}(t))
\tilde{z}_{\{i_1, \dots, i_n\}} \nonumber \\
&   & + \sum_{k = 0}^{n-1} \sum_{\{j_1, \dots, j_k\} \subset
\{i_1, \dots, i_n\}} \kappa_{\{j_1, \dots, j_k\}}
\tilde{z}_{\{j_1, \dots, j_k\}} \times \nonumber \\
&   & \prod_{i \in \{i_1, \dots, i_n\}/\{j_1, \dots, j_k\}} (1 -
e^{-\alpha_i \mu/2})
\end{eqnarray}

Finally, another subject for future work is the incorporation of
repair into our network model.  In \cite{REPFULL, REPFIRST} it was
assumed that only one gene controlled repair.  By assuming that
several genes control repair, then, in analogy with fitness, we
hypothesize that instead of a single ``repair catastrophe''
\cite{REPFULL, REPFIRST}, we obtain a series of localization to
delocalization transitions over the repair gene sequence spaces, a
``repair cascade.''

\begin{acknowledgments}

This research was supported by the National Institutes of Health.
The authors would like to thank Eric J. Deeds for helpful 
discussions.

\end{acknowledgments}

\begin{appendix}

\begin{widetext}

\section{Derivation of the Reduced System of Equations}

In this appendix, we derive Eq. (13) from Eq. (11).  To this end,
define,
\begin{equation}
z_{\{i_1, \dots, i_n\}} = \sum_{l_{i_1} = 0}^{\infty}
\cdot \dots \cdot \sum_{l_{i_n} = 0}^{\infty}
{z_{l_{i_1} {\bf e}_{i_1} + \dots + l_{i_n} {\bf e}_{i_n}}}
\end{equation}
We then have that,
\begin{eqnarray}
\frac{d z_{\{i_1, \dots, i_n\}}}{dt} & = & \sum_{l_{i_1} =
0}^{\infty} \cdot \dots \cdot \sum_{l_{i_n} = 0}^{\infty}
(e^{-\mu} \sum_{l_{i_1}' = 0}^{l_{i_1}} \cdot
\dots \cdot \sum_{l_{i_n}' = 0}^{l_{i_n}} \times \nonumber \\
&   &
\frac{\kappa_{(l_{i_1} - l_{i_1}') {\bf e}_{i_1} + \dots
+ (l_{i_n} - l_{i_n}') {\bf e}_{i_n}}}{l_{i_1}'! \cdot \dots
\cdot l_{i_n}'!} (\alpha_{i_1} \mu)^{l_{i_1}'} \cdot \dots
\cdot (\alpha_{i_n} \mu)^{l_{i_n}'} z_{(l_{i_1} - l_{i_1}')
{\bf e}_{i_1} + \dots + (l_{i_n} - l_{i_n}') {\bf e}_{i_n}}
\nonumber \\
&   &
- \bar{\kappa}(t) z_{l_{i_1} {\bf e}_{i_1} + \dots + l_{i_n}
{\bf e}_{i_n}}) \nonumber \\
& = &
e^{-\mu}
\sum_{l_{i_1}' = 0}^{\infty} \cdot \dots \cdot
\sum_{l_{i_n}' = 0}^{\infty}
{\frac{1}{l_{i_1}'! \cdot \dots \cdot l_{i_n}'!}
(\alpha_{i_1} \mu)^{l_{i_1}'} \cdot \dots
\cdot (\alpha_{i_n} \mu)^{l_{i_n}'}} \times \nonumber \\
&   &
\sum_{l_{i_1} = l_{i_1}'}^{\infty} \cdot \dots \cdot
\sum_{l_{i_n} = l_{i_n}'}^{\infty}
{\kappa_{(l_{i_1} - l_{i_1}') {\bf e}_{i_1} + \dots +
(l_{i_n} - l_{i_n}') {\bf e}_{i_n}}
z_{(l_{i_1} - l_{i_1}') {\bf e}_{i_1} + \dots +
(l_{i_n} - l_{i_n}') {\bf e}_{i_n}}} -
\bar{\kappa}(t) z_{\{i_1, \dots, i_n\}}
\nonumber \\
& = &
e^{-(1 - \alpha_{i_1} - \dots - \alpha_{i_n})\mu}
\sum_{k_{i_1} = 0}^{\infty} \cdot \dots \cdot
\sum_{k_{i_n} = 0}^{\infty}
{\kappa_{k_{i_1} {\bf e}_{i_1} + \dots +
k_{i_n} {\bf e}_{i_n}} z_{k_{i_1} {\bf e}_{i_1} +
\dots + k_{i_n} {\bf e}_{i_n}}}
- \bar{\kappa}(t) z_{\{i_1, \dots, i_n\}}
\nonumber \\
& = &
e^{-(1 - \alpha_{i_1} - \dots - \alpha_{i_n})\mu}
\sum_{k = 0}^{n} \sum_{\{j_1, \dots, j_k\} \subseteq
\{i_1, \dots, i_n\}}
{\kappa_{\{j_1, \dots, j_k\}} \tilde{z}_{\{j_1, \dots, j_k\}}}
- \bar{\kappa}(t) z_{\{i_1, \dots, i_n\}}
\end{eqnarray}

We now claim that,
\begin{equation}
\tilde{z}_{\{i_1, \dots, i_n\}} =
\sum_{k = 0}^{n} (-1)^{n-k} \sum_{\{j_1, \dots, j_k\} \subseteq
\{i_1, \dots, i_n\}}{z_{\{j_1, \dots, j_k\}}}
\end{equation}
This can be proved by induction.  For $ n = 0 $ this statement is clearly
true, since $ z_{\emptyset} = \tilde{z}_{\emptyset} $.  Suppose then, that
for some $ n \geq 0 $, the statement is true for all $ 0 \leq m \leq n $.
Then we have,
\begin{eqnarray}
z_{\{i_1, \dots, i_{n+1}\}}
& = &
\sum_{k = 0}^{n+1} \sum_{\{j_1, \dots, j_k\} \subseteq
\{i_1, \dots, i_{n+1}\}}{\tilde{z}_{\{j_1, \dots, j_k\}}} \nonumber \\
& = & \tilde{z}_{\{i_1, \dots, i_{n+1}\}} \nonumber \\
&   & + \sum_{k = 0}^{n} \sum_{\{j_1, \dots, j_k\} \subseteq
\{i_1, \dots, i_{n+1}\}} {\tilde{z}_{\{j_1, \dots, j_k\}}}
\nonumber \\
\end{eqnarray}
and so,
\begin{eqnarray}
\tilde{z}_{\{i_1, \dots, i_{n+1}\}} & = & z_{\{i_1, \dots,
i_{n+1}\}}
\nonumber \\
&   & - \sum_{k = 0}^{n} \sum_{\{j_1, \dots, j_k\} \subseteq
\{i_1, \dots, i_{n+1}\}} \sum_{l = 0}^{k} (-1)^{k-l} \times
\nonumber
\\
&   & \sum_{\{j_1', \dots, j_l'\} \subseteq \{j_1, \dots, j_k\}}
{z_{\{j_1', \dots, j_k'\}}}
\end{eqnarray}
Now, for each set $ \{j_1, \dots, j_k\} $ appearing in the sum, a given
subset $ \{j_1', \dots, j_l'\} $ occurs only once.  The $ k $-element sets
$ \{j_1, \dots, j_k\} $ which contain $ \{j_1', \dots, j_l'\} $ as a
subset must be of the form $ \{j_1', \dots, j_l'\} \cup
\{j_1'', \dots, j_{k-l}''\} $, where $ \{j_1'', \dots, j_{k-l}''\} \subseteq
\{i_1, \dots, i_{n+1}\}/\{j_1', \dots, j_l'\} $.  Therefore, there are $
{n+1-l}\choose{k-l} $ distinct $ k $-element sets which contain
$ \{j_1,', \dots, j_l'\} $.  Rearranging the sum, we obtain,
\begin{eqnarray}
\tilde{z}_{\{i_1, \dots, i_{n+1}\}} & = & z_{\{i_1, \dots,
i_{n+1}\}} \nonumber \\
&   & - \sum_{l = 0}^{n} \sum_{\{j_1, \dots, j_l\} \subseteq
\{i_1, \dots, i_{n+1}\}} z_{\{j_1, \dots, j_l\}} \times \nonumber
\\
&   &
\sum_{k = l}^{n}
(-1)^{k - l} {{n+1-l}\choose{k-l}} \nonumber \\
& = & z_{\{i_1, \dots, i_{n+1}\}} \nonumber \\
&   &
- \sum_{l = 0}^{n} \sum_{\{j_1, \dots, j_l\} \subseteq \{i_1,
\dots, i_{n+1}\}} z_{\{j_1, \dots, j_l\}}
(-(-1)^{n+1-l}) \nonumber \\
& = &
\sum_{l = 0}^{n+1}
(-1)^{n+1-l}
\sum_{\{j_1, \dots, j_l\} \subseteq \{i_1, \dots, i_{n+1}\}}
z_{\{j_1, \dots, j_l\}}
\end{eqnarray}
This completes the induction step, and proves the claim.

We are almost ready to derive the expression for $ d \tilde{z}_{\{i_1, \dots,
i_n\}}/dt $.  Before doing so, we state the following identity, which
we will need in our calculation:
\begin{equation}
\prod_{i = 1}^{n} (1 - \alpha_i) = \sum_{k = 0}^{n} (-1)^k
\sum_{\{i_1, \dots, i_k\} \subseteq \{1, \dots, n\}} \alpha_{i_1}
\cdot \dots \cdot \alpha_{i_k}
\end{equation}
We now have,
\begin{eqnarray}
\frac{d \tilde{z}_{\{i_1, \dots, i_n\}}}{dt}
& = &
\sum_{k = 0}^{n} (-1)^{n-k}
\sum_{\{j_1, \dots, j_k\} \subseteq \{i_1, \dots, i_n\}}
\frac{d z_{\{j_1, \dots, j_k\}}}{dt} \nonumber \\
& = &
\sum_{k = 0}^{n} (-1)^{n-k}
\sum_{\{j_1, \dots, j_k\} \subseteq \{i_1, \dots, i_n\}}
(e^{-(1 - \alpha_{j_1} - \dots - \alpha_{j_k})\mu}
\sum_{l = 0}^{k}
\sum_{\{j_1', \dots, j_l'\} \subseteq \{j_1, \dots, j_k\}}
{\kappa_{\{j_1', \dots, j_l'\}} \tilde{z}_{\{j_1', \dots, j_l'\}}}
\nonumber \\
&   &
- \bar{\kappa}(t) z_{\{j_1, \dots, j_k\}}) \nonumber \\
& = &
\sum_{l = 0}^{n}
\sum_{\{j_1, \dots, j_l\} \subseteq \{i_1, \dots, i_n\}}
\kappa_{\{j_1, \dots, j_l\}} \tilde{z}_{\{j_1, \dots, j_l\}}
\times \nonumber \\
&   &
\sum_{k = l}^{n} (-1)^{n-k}
\sum_{\{j_1', \dots, j_{k-l}'\} \subseteq
\{i_1, \dots, i_n\}/\{j_1, \dots, j_l\}}
e^{-(1 - \alpha_{j_1} - \dots - \alpha_{j_l} - \alpha_{j_1}' - \dots
- \alpha_{j_{k-l}'})\mu} \nonumber \\
&   &
- \bar{\kappa}(t) \tilde{z}_{\{i_1, \dots, i_n\}} \nonumber \\
& = &
\sum_{l = 0}^{n}
\sum_{\{j_1, \dots, j_l\} \subseteq \{i_1, \dots, i_n\}}
\kappa_{\{j_1, \dots, j_l\}} \tilde{z}_{\{j_1, \dots, j_l\}}
\times \nonumber \\
&   & e^{-(1 - \alpha_{j_1} - \dots - \alpha_{j_l})\mu} \sum_{k-l
= 0}^{n-l} (-1)^{n-l} (-1)^{k-l} \sum_{\{j_1', \dots, j_{k-l}'\}
\subseteq \{i_1, \dots, i_n\}/\{j_1, \dots, j_l\}}
e^{\alpha_{j_1'} \mu} \cdot \dots \cdot e^{\alpha_{j_{k-l}'}\mu}
\nonumber \\
&   &
- \bar{\kappa}(t) \tilde{z}_{\{i_1, \dots, i_n\}} \nonumber \\
& = &
\sum_{l = 0}^{n}
\sum_{\{j_1, \dots, j_l\} \subseteq \{i_1, \dots, i_n\}}
\kappa_{\{j_1, \dots, j_l\}} \tilde{z}_{\{j_1, \dots, j_l\}} \times
\nonumber \\
&   & (-1)^{n-l} e^{-(1 - \alpha_{j_1} - \dots - \alpha_{j_l})\mu}
\prod_{i \in \{i_1, \dots, i_n\}/\{j_1, \dots, j_l\}}
(1 - e^{\alpha_i \mu}) \nonumber \\
&   &
- \bar{\kappa}(t) \tilde{z}_{\{i_1, \dots, i_n\}} \nonumber \\
& = & e^{-(1 - \alpha_{i_1} - \dots - \alpha_{i_n})\mu} \sum_{l =
0}^{n} \sum_{\{j_1, \dots, j_l\} \subseteq \{i_1, \dots, i_n\}}
\kappa_{\{j_1, \dots, j_l\}} \tilde{z}_{\{j_1, \dots, j_l\}}
\prod_{i \in \{i_1, \dots, i_n\}/\{j_1, \dots, j_l\}}
(1 - e^{-\alpha_i \mu}) \nonumber \\
&   &
- \bar{\kappa}(t) \tilde{z}_{\{i_1, \dots, i_n\}}
\end{eqnarray}
which is exactly Eq. (13).

\section{Derivation of the Localization Lengths}

In this section we derive the expression for $ d \tilde{\langle
l_i \rangle}_{\{i_1, \dots, i_n\}}/dt $. We have,
\begin{eqnarray}
\frac{d \langle l_i\rangle_{\{i_1, \dots, i_n\}}}{dt}
& = &
\sum_{l_{i_1} = 0}^{\infty} \cdot \dots \cdot \sum_{l_{i_n} = 0}^{\infty}
\sum_{l_i = 0}^{\infty}
l_i (e^{-\mu}
\sum_{l_{i_1}' = 0}^{l_{i_1}} \cdot \dots \cdot
\sum_{l_{i_n}' = 0}^{l_{i_n}}
\sum_{l_i' = 0}^{l_i}
\frac{\kappa_{(l_{i_1} - l_{i_1}') {\bf e}_{i_1} + \dots +
               (l_{i_n} - l_{i_n}') {\bf e}_{i_n} +
               (l_i - l_i') {\bf e}_i}}
      {l_{i_1}'! \cdot \dots \cdot l_{i_n}'! l_i'!} \times \nonumber \\
&   &
      (\alpha_{i_1} \mu)^{l_{i_1}'} \cdot \dots \cdot
      (\alpha_{i_n} \mu)^{l_{i_n}'}
      (\alpha_i \mu)^{l_i'}
      z_{(l_{i_1} - l_{i_1}') {\bf e}_{i_1} + \dots +
         (l_{i_n} - l_{i_n}') {\bf e}_{i_n} +
         (l_i - l_i') {\bf e}_i} \nonumber \\
&   &
- \bar{\kappa}(t) z_{l_{i_1} {\bf e}_{i_1} + \dots + l_{i_n} {\bf e}_{i_n} +
                   l_i {\bf e}_i}) \nonumber \\
& = &
e^{-\mu}
\sum_{l_{i_1}' = 0}^{\infty} \cdot \dots \cdot
\sum_{l_{i_n}' = 0}^{\infty}
\sum_{l_i' = 0}^{\infty}
\frac{(\alpha_{i_1} \mu)^{l_{i_1}'} \cdot \dots \cdot
      (\alpha_{i_n} \mu)^{l_{i_n}'}
      (\alpha_i \mu)^{l_i'}}
     {l_{i_1}'! \cdot \dots \cdot l_{i_n}'! l_i'!} \times \nonumber \\
&   &
\sum_{k_{i_1} = 0}^{\infty} \cdot \dots \cdot
\sum_{k_{i_n} = 0}^{\infty}
\sum_{k_i = 0}^{\infty}
(k_i + l_i') \kappa_{k_{i_1} {\bf e}_{i_1} + \dots + k_{i_n} {\bf e}_{i_n} +
                     k_i {\bf e}_i}
             z_{k_{i_1} {\bf e}_{i_1} + \dots + k_{i_n} {\bf e}_{i_n} +
                k_i {\bf e}_i} \nonumber \\
&   &
- \bar{\kappa}(t) \langle l_i \rangle_{\{i_1, \dots, i_n\}} \nonumber \\
& = &
e^{-(1 - \alpha_{i_1} - \dots - \alpha_{i_n} - \alpha_i)\mu}
(\sum_{k = 0}^{n}\sum_{\{j_1, \dots, j_k\} \subseteq \{i_1, \dots, i_n\}}
{\kappa_{\{j_1, \dots, j_k, i\}}
\tilde{\langle l_i \rangle}_{\{j_1, \dots, j_k\}}} \nonumber \\
&   &
+ \alpha_i \mu
\sum_{k = 0}^{n+1}\sum_{\{j_1, \dots, j_k\} \subseteq \{i_1, \dots, i_n, i\}}
{\kappa_{\{j_1, \dots, j_k\}}
\tilde{z}_{\{j_1, \dots, j_k\}}})
- \bar{\kappa}(t) \langle l_i \rangle_{\{i_1, \dots, i_n\}} \nonumber \\
& = &
e^{-(1 - \alpha_{i_1} - \dots - \alpha_{i_n} - \alpha_i)\mu}
\sum_{k = 0}^{n} \sum_{\{j_1, \dots, j_k\} \subseteq \{i_1, \dots, i_n\}}
\times \nonumber \\
&   &
(\kappa_{\{j_1, \dots, j_k, i\}}
 \tilde{\langle l_i \rangle}_{\{j_1, \dots, j_k\}} +
 \alpha_i \mu \kappa_{\{j_1, \dots, j_k\}}
              \tilde{z}_{\{j_1, \dots, j_k\}} +
 \alpha_i \mu \kappa_{\{j_1, \dots, j_k, i\}}
              \tilde{z}_{\{j_1, \dots, j_k, i\}}) \nonumber \\
&   &
- \bar{\kappa}(t) \langle l_i \rangle_{\{i_1, \dots, i_n\}}
\end{eqnarray}

We therefore have that,
\begin{eqnarray}
\frac{d \tilde{\langle l_i\rangle}_{\{i_1, \dots, i_n\}}}{dt}
& = &
\sum_{k = 0}^{n} (-1)^{n-k}
\sum_{\{j_1, \dots, j_k\} \subseteq \{i_1, \dots, i_n\}}
{\frac{d \langle l_i\rangle_{\{j_1, \dots, j_k\}}}{dt}} \nonumber \\
& = &
\sum_{k = 0}^{n} (-1)^{n-k}
\sum_{\{j_1, \dots, j_k\} \subseteq \{i_1, \dots, i_n\}}
[e^{-(1 - \alpha_{j_1} - \dots - \alpha_{j_k} - \alpha_i)\mu}
\sum_{l = 0}^{k}\sum_{\{j_1', \dots, j_l'\} \subseteq \{j_1, \dots, j_k\}}
\times \nonumber \\
&   &
(\kappa_{\{j_1', \dots, j_l', i\}}
 \tilde{\langle l_i\rangle}_{\{j_1', \dots, j_l'\}} +
 \alpha_i \mu \kappa_{\{j_1', \dots, j_l'\}}
              \tilde{z}_{\{j_1', \dots, j_l'\}} +
 \alpha_i \mu \kappa_{\{j_1', \dots, j_l', i\}}
              \tilde{z}_{\{j_1', \dots, j_l', i\}}) \nonumber \\
&   &
- \bar{\kappa}(t) \langle l_i\rangle_{\{j_1, \dots, j_k\}}] \nonumber \\
& = &
\sum_{l = 0}^{n} \sum_{\{j_1', \dots, j_l'\} \subseteq \{i_1, \dots, i_n\}}
(-1)^{n-l} e^{-(1 - \alpha_{j_1'} - \dots - \alpha_{j_l}' - \alpha_i) \mu}
\times \nonumber \\
&   &
(\kappa_{\{j_1', \dots, j_l', i\}}
 \tilde{\langle l_i\rangle}_{\{j_1', \dots, j_l'\}} +
 \alpha_i \mu \kappa_{\{j_1', \dots, j_l'\}}
              \tilde{z}_{\{j_1', \dots, j_l'\}} +
 \alpha_i \mu \kappa_{\{j_1', \dots, j_l', i\}}
              \tilde{z}_{\{j_1', \dots, j_l', i\}}) \times \nonumber \\
&   & \sum_{k-l = 0}^{n-l} (-1)^{k-l} \sum_{\{j_1, \dots,
j_{k-l}\} \subseteq \{i_1, \dots, i_n\}/\{j_1', \dots, j_l'\}}
e^{\alpha_{j_1} \mu} \cdot \dots \cdot e^{\alpha_{j_{k-l}} \mu}
\nonumber \\
&   &
- \bar{\kappa}(t) \tilde{\langle l_i\rangle}_{\{i_1, \dots, i_n\}}
\nonumber \\
& = &
e^{-(1 - \alpha_{i_1} - \dots - \alpha_{i_n} - \alpha_i)\mu}
\sum_{k = 0}^{n} \sum_{\{j_1, \dots, j_k\} \subseteq \{i_1, \dots, i_n\}}
\times \nonumber \\
&   &
(\kappa_{\{j_1, \dots, j_k, i\}}
 \tilde{\langle l_i\rangle}_{\{j_1, \dots, j_k\}} +
 \alpha_i \mu \kappa_{\{j_1, \dots, j_k\}}
              \tilde{z}_{\{j_1, \dots, j_k\}} +
 \alpha_i \mu \kappa_{\{j_1, \dots, j_k, i\}}
              \tilde{z}_{\{j_1, \dots, j_k, i\}}) \times \nonumber \\
&   & \prod_{j \in \{i_1, \dots, i_n\}/\{j_1, \dots, j_k\}} (1 -
e^{-\alpha_j \mu})
\nonumber \\
&   &
- \bar{\kappa}(t) \tilde{\langle l_i\rangle}_{\{i_1, \dots, i_n\}}
\end{eqnarray}
which is exactly the expression in Eq. (25).

\section{Numerical Details}

The finite sequence length equations, given by Eq. (11), may be
expressed in vector form,
\begin{equation}
\frac{d \vec{z}}{dt} = {\bf B} \vec{z} - (\vec{\kappa} \cdot
\vec{z}) \vec{z}
\end{equation}
At equilibrium, we therefore have that,
\begin{equation}
\vec{z} = \frac{1}{\vec{\kappa} \cdot \vec{z}} {\bf B} \vec{z}
\end{equation}
The equilibrium solution may be found using fixed-point iteration,
via the equation,
\begin{equation}
\vec{z}_{n+1} = \frac{1}{\vec{\kappa} \cdot \vec{z}_n} {\bf B}
\vec{z}_n
\end{equation}
The iterations are stopped when the $ z_n $ stop changing.  This
is determined by introducing a cutoff parameter $ \delta $, and
stop iterating when the fractional change of each of the
components after $ N_\epsilon $ iterations is smaller than $
\delta $.  $ N_\epsilon $ is chosen to be sufficiently large so
that, on average, each base mutates at least once after $
N_\epsilon $ iterations.  Thus, we choose $ N_\epsilon =
1/\epsilon $.

What this method does is account for the fact that equilibration
takes longer for smaller values of $ \epsilon $.  This means that
the smaller the value of $ \epsilon $, the more times it is
necessary to iterate before comparing the changes in the $
\vec{z}_n $.  For our two-gene simulation, we took $ \delta =
10^{-4} $, and $ \vec{z}_{0} = (1, 1) $.  We chose this initial
condition to show that, even though backmutations may become small
at large sequence lengths, they still strongly affect the
equilibrium solution.  By iterating a sufficient number of times,
the cumulative effect of the backmutations becomes sufficiently
large to lead to a unique equilibrium solution, independent of the
initial condition.

\end{widetext}

\end{appendix}

\end{document}